\documentclass[fleqn,usenatbib]{mnras}

\usepackage[T1]{fontenc}

\DeclareRobustCommand{\VAN}[3]{#2}
\let\VANthebibliography\thebibliography
\def\thebibliography{\DeclareRobustCommand{\VAN}[3]{##3}\VANthebibliography}

\usepackage{rotating}
\usepackage{tabularx}
\usepackage{multirow}
\usepackage{multicol}
\usepackage{geometry}
\usepackage{newtxtext,newtxmath}
\usepackage{xcolor}

\usepackage{graphicx}	
\usepackage{amsmath}	
\usepackage{ae,aecompl}
\usepackage{lscape}

\newcommand{\mdwarf}{M-dwarf}
\newcommand{\mdwarfs}{M-dwarfs}
\newcommand{\tic}{TICv8}
\newcommand{\siglc}{$\sigma_{\rm LC}$}
\newcommand{\focc}{$f_{\rm occ}$}

\newcommand{\kepler}{{\it Kepler}}

\newcommand{\kms}{km\,s$^{-1}$}

\newcommand{\mpl}{\mbox{$M_{\rm P}$}}
\newcommand{\rpl}{\mbox{$R_{\rm P}$}}

\newcommand{\mstar}{\mbox{$M_{*}$}}
\newcommand{\rstar}{\mbox{$R_{*}$}}
\newcommand{\rhostar}{\mbox{$\rho_{*}$}}
\newcommand{\mjup}{\mbox{$M_{\rm J}$}}
\newcommand{\rjup}{\mbox{$R_{\rm J}$}}

\newcommand{\mearth}{\mbox{$M_{\oplus}$}}
\newcommand{\rearth}{\mbox{$R_{\oplus}$}}
\newcommand{\msun}{\mbox{$M_{\odot}$}}
\newcommand{\rsun}{\mbox{$R_{\odot}$}}
\newcommand{\gccc}{g\,cm$^{-3}$}

\newcommand{\teff}{$T_{\rm eff}$}

\newcommand{\logg}{$\log g$}
\newcommand{\tc}{$T_C$}
\newcommand{\rprs}{\mbox{R$_{P}/$R$_{*}$}}

\newcommand{\Nfoccfull}{\mbox{$0.194 \pm 0.072$\,\%}}%
\newcommand{\Nfoccfullmax}{\mbox{$0.267 \pm 0.079$\,\%}}%
\newcommand{\Nfoccfullmin}{\mbox{$0.139 \pm 0.066$\,\%}}%

\newcommand{\Nfoccfullnopc}{\mbox{$0.194 \pm 0.072$}}%
\newcommand{\Nfoccfullmaxnopc}{\mbox{$0.267 \pm 0.079$}}%
\newcommand{\Nfoccfullminnopc}{\mbox{$0.139 \pm 0.066$}}%

\newcommand{\Nfocclowmass}{\mbox{$0.137 \pm 0.097$}\,\%}%
\newcommand{\Nfoccmedmass}{\mbox{$0.108 \pm 0.083$}\,\%}%
\newcommand{\Nfocchighmass}{\mbox{$0.29 \pm 0.15$}\,\%}%

\newcommand{\Nfocclowmassnopc}{\mbox{$0.137 \pm 0.097$}}%
\newcommand{\Nfoccmedmassnopc}{\mbox{$0.108 \pm 0.083$}}%
\newcommand{\Nfocchighmassnopc}{\mbox{$0.29 \pm 0.15$}}%

\newcommand{\Nfoccbelowburn}{\mbox{$0.134 \pm 0.069$\,\%}}%

\newcommand{\Nfoccbelowburnnopc}{\mbox{$0.134 \pm 0.069$}}%
\newcommand{\Nfoccbelowburnmaxnopc}{\mbox{$0.249 \pm 0.090$}}%
\newcommand{\Nfoccbelowburnminnopc}{\mbox{$0.074\pm 0.053$}}%

\usepackage{newtxtext,newtxmath}

\title[Giant planets orbiting low-mass stars]{The occurrence rate of giant planets orbiting low-mass stars with TESS}

\author[E. M. Bryant et al.]{
\parbox{\textwidth}{
Edward M. Bryant,$^{1, 2, 3}$\thanks{E-mail:\href{edward.bryant@ucl.ac.uk}{edward.bryant@ucl.ac.uk}}
Daniel Bayliss,$^{2, 3}$
Vincent Van Eylen$^{1}$
}
\\
$^{1}$Mullard Space Science Laboratory, University College London, Holmbury St Mary, Dorking, Surrey, RH5 6NT, UK\\
$^{2}$Dept.\ of Physics, University of Warwick, Gibbet Hill Road, Coventry CV4 7AL, UK\\
$^{3}$Centre for Exoplanets and Habitability, University of Warwick, Gibbet Hill Road, Coventry CV4 7AL, UK\\
}

\date{Accepted 23 February 2023. Received 02 February 2023; in original form 02 December 2022.}

\pubyear{2023}

\begin{document}
\label{firstpage}
\pagerange{\pageref{firstpage}--\pageref{lastpage}}
\maketitle

\begin{abstract}
We present a systematic search for transiting giant planets ($0.6 \rjup \leq \rpl \leq 2.0 \rjup$) orbiting nearby low-mass stars ($\mstar \leq 0.71 \msun$). The formation of giant planets around low-mass stars is predicted to be rare by the core-accretion planet formation theory. We search 91,306 low-mass stars in the TESS 30\,minute cadence photometry detecting fifteen giant planet candidates, including seven new planet candidates which were not known planets or identified as TOIs prior to our search. Our candidates present an exciting opportunity to improve our knowledge of the giant planet population around the lowest mass stars. We perform planet injection-recovery simulations and find that our pipeline has a high detection efficiency across the majority of our targeted parameter space. We measure the occurrence rates of giant planets with host stars in different stellar mass ranges spanning our full sample. We find occurrence rates of \Nfocclowmass\ (0.088 - 0.26\,\msun), \Nfoccmedmass\ (0.26 - 0.42\,\msun), and \Nfocchighmass\ (0.42 - 0.71\,\msun). For our full sample (0.088 - 0.71\,\msun) we find a giant planet occurrence rate of \Nfoccfull. We have measured for the first time the occurrence rate for giant planets orbiting stars with $\mstar \leq 0.4\,\msun$ and we demonstrate this occurrence rate to be non-zero. This result contradicts currently accepted planet formation models and we discuss some possibilities for how these planets could have formed.
\end{abstract}



\section{Introduction}
Since the detection of 51-Pegasi b \citep{mayorqueloz199551peg} revealed the existence of giant planets on short orbits ($P \lesssim 10$\,d), known as hot Jupiters, the prevalence of these planets has been a question of great interest to the community. As such, there have been numerous studies that have focused on measuring the occurrence rate of these planets. Different planet formation mechanisms should leave signatures in the occurrence rates of the different populations of planets \citep[eg.][]{emsenhuber2021bernngpps, emsenhuber2021bernngpps2, schlecker2021bernngpps}. For example, it has been demonstrated that gas giant planets are found predominantly around high metallicity host stars \citep{fischervalenti2005pmc, osborn2020pmc}. It is believed that the metallicity of a star is inherited from its primordial cloud and thus shared by its protoplanetary disk \citep{fischer2005planetmetallicity}, and it has been shown that gas giant planets from through core-accretion more efficiently in these metal-rich protoplanetary disks \citep[e.g.][]{idalin2004gasgiantformation, thommes2008giantformation}. As such, this giant planet metallicity correlation has been interpreted as evidence that giant planets predominantly form through the core-accretion mechanism \citep[e.g. ][]{johnson2010occurrencemetallicity}. 

\begin{figure}
    \centering
    \includegraphics[width=0.95\columnwidth]{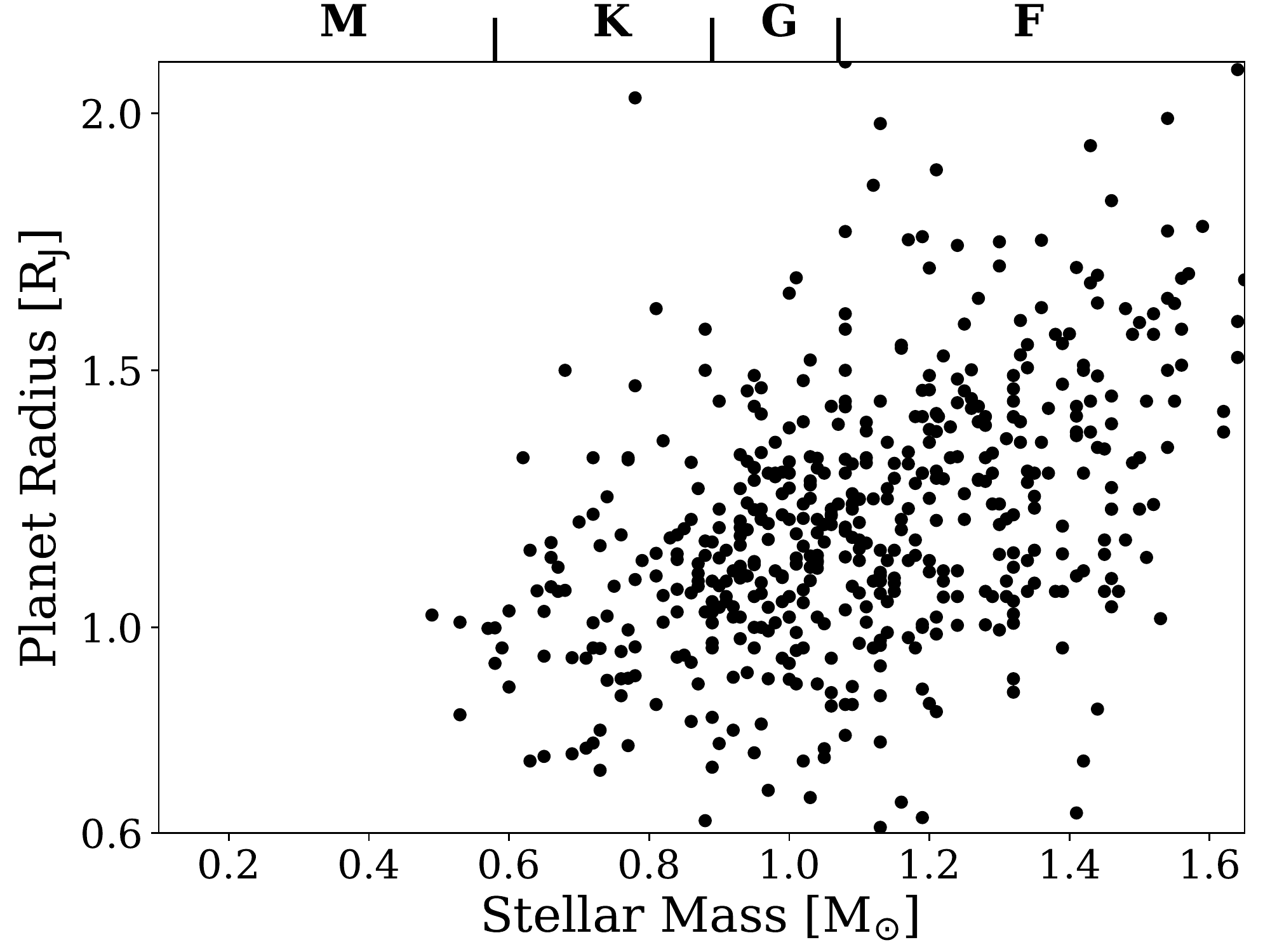}
    \caption[Radius of known transiting giant planets as a function of host star mass]{Known transiting exoplanets as a function of the mass of the host star (data accessed from the \href{https://exoplanetarchive.ipac.caltech.edu/index.html}{NASA Exoplanet Archive} on 2022 November 16). The sample shown is made up of planets with masses between 0.1 and 13\,\mjup\ and radii larger than 0.6\,\rjup.}
    \label{fig:ms_mpl}
\end{figure}

Over the past two decades, the number of known exoplanets has exploded, particularly as a result of the \kepler\ mission \citep{borucki2010kepler}, resulting in huge advances in our knowledge of hot Jupiter occurrence rates. Using early \kepler\ results for a sample of G- and K-type stars (\teff$ = 4100 - 6100 $ \,K and $\log g = 4.0 - 4.9$), \citet{howard2012kepleroccurrence} derived a frequency of $0.4 \pm 0.1$\% for planets with $\rpl \geq 8$\,\rearth\ and $P < 10$\,d. Improving upon the estimates of the rate of false positives in the \kepler\ data, \citet{fressin2013kepleroccurrences} refined the \kepler\ occurrence rate estimate to $0.43 \pm 0.05$\% for planets with $6\,\rearth \leq \rpl \leq 22\,\rearth$ and $0.8 \leq P \leq 10$\,d.

Despite providing a deep insight into the occurrence rates of planets around FGK stars, \kepler\ has not been able to provide similar results for the occurrence rate of giant planets orbiting low-mass stars.  \kepler\ only monitored roughly 4000 stars with \teff$ \leq 4000\,$K \citep{dressingcharbonneau2013mdwarfkepleroccrates}. With the detection of 91 transiting planet candidates with $\rpl \leq 4.0\,\rearth$ around these stars \citet{dressingcharbonneau2013mdwarfkepleroccrates} could strongly constrain the occurrence rates of small radii planets with low-mass host stars and periods under 50\,days to be $90^{+4}_{-3}$\,\%. Unfortunately the low number of M-dwarfs monitored by \kepler\ lead to the discovery of just three planet candidates with $\rpl > 0.6\rjup$ \citep{dressingcharbonneau2013mdwarfkepleroccrates}. Based on their three candidates \citet{dressingcharbonneau2013mdwarfkepleroccrates} derived an occurrence rate for giant planets on short orbital periods ($P \leq 10\,d$) of $1^{+0.8}_{-0.2}$\,\%., However, of these three candidates just one -- Kepler-45\,b \citep{johnson2012kepler45} -- has been confirmed as a giant planet. We note that the other two have since been shown to be a $1.97\,\rearth$ planet \citep[Kepler-1124\,b;][]{morton2016kep1124, cloutiermenou2020kep1124} and an \mdwarf-white dwarf binary system \citep[KOI-256;][]{muirhead2013koi256}.  If we instead consider now the fact that, from these three candidates, just Kepler-45\,b is real, then from the \citet{dressingcharbonneau2013mdwarfkepleroccrates} results we would obtain an occurrence rate of $0.3^{+0.44}_{0.1}$\,\%, which is statistically consistent with the hot Jupiter occurrence rate measured by \kepler\ \citep[e.g.][]{fressin2013kepleroccurrences, petigura2018cksoccurrencerates}. Therefore, the \kepler\ results alone are unable to conclude that the giant planet occurrence rate for low-mass stars is significantly lower than for solar-like stars.

Determining the frequency of giant planets around low-mass stars will act as a key test for planet formation theories. It has been shown that giant planets form less readily through core accretion around low-mass stars than Sun-like stars \citep{laughlin2004coreaccretion, idalin2005planetformation}. Theoretical population synthesis models from \citet{burn2021ngpps} predict that the occurrence rate of giant planets ($\mpl \geq 30$\,\mearth) with 0.7\,\msun\ host stars would be approximately half that for a 1\,\msun\ host star. For 0.5\,\msun\ host stars they predict an occurrence rate roughly 13\% that of a 1\,\msun\ host star, and for host stars with $\mstar < 0.5$\,\msun\ their results find giant planet occurrence rates of zero, predicting that the formation of giant planets through core-accretion is impossible for these low-mass host stars. 

Recent discoveries of hot Jupiter planets orbiting \mdwarfs\ \citep[eg.][]{hartman2015hats6, bayliss2018ngts1,bakos2020hats71} have shown that giant planets can form around low-mass stars. However, despite these recent discoveries, there are still very few known hot Jupiters transiting low-mass stars and the overall population is still not well understood. From the 536 transiting giant exoplanets known to date, only sixteen orbit a star with a stellar mass less than 0.65\,\msun, and only one -- HATS-71\,b \citep{bakos2020hats71} -- orbits a star with a stellar mass less than 0.5\,\msun\footnote{Data accessed from \href{https://exoplanetarchive.ipac.caltech.edu/index.html}{NASA Exoplanet Archive} on 2022 November 16.} (see Figure~\ref{fig:ms_mpl}). The true impact of these discoveries on our knowledge of planet formation cannot be determined without a robust measurement of the occurrence rate of these systems.

Attempts have been made to measure the occurrence rates for giant planets with low-mass host stars, but the majority have not yielded conclusive results. The Pan Planets survey \citep{obermeier2016panplanets} placed a 95\% confidence level upper limit of 0.34\% on the occurrence rates for hot Jupiters with host stars with an effective temperature \teff$ < 3900$\,K and surface gravity $\log g > 4$, concluding that they could not confirm a significant difference in the occurrence rate of hot Jupiters compared to those with solar-type host stars. More recently, \citet{sabotta2021carmenesoccurrence} attempted to derive planet occurrence rates around stars with masses $\mstar \leq 0.7$\,\msun using data from CARMENES. This study suffers from a small stellar sample size, including data for only 71 low-mass stars, and did not detect any giant planets with $P \leq 10$\,d. Therefore, they could only place an upper limit of 3\% on giant planet frequency orbiting low-mass stars. \citet{gan2022tesshjoccurrence} have used the TESS photometry to constrain the occurrence rate of hot Jupiters ($7 \rearth \leq \rpl \leq 2 \rjup$; $0.8 \leq P \leq 10$\,d) around early M-dwarfs ($0.45 - 0.65 \msun$) to $0.27\pm0.09$\%. 

In this paper we present a systematic search for transiting gas giant planets ($0.6 \rjup \leq \rpl \leq 2.0 \rjup$) orbiting nearby ($d \leq 100$\,pc), low-mass ($0.088\,\msun \leq \mstar \leq 0.71\,\msun$) stars in the TESS 30\,minute cadence light curves \citep{caldwell2020tessspoc}. This search has been motivated and designed to measure the occurrence rates of giant planets on close-in orbits ($1\,{\rm d} \leq P \leq 10\,{\rm d}$) around low-mass stars.

Through this study we have measured giant planet occurrence rates for lower-mass host stars than previous studies. In addition to measuring these occurrence rates, our search has also yielded the discovery of seven new giant planet candidates which were not previously known, either as confirmed planets or TOI planet candidates, for which follow-up work to confirm their true natures is underway. We present our new candidates in this work and we interpret our results in the context of giant planet formation.

We detail our sample selection in Section~\ref{sec:lmstar_sample}. We present our transit search in Section~\ref{sec:transit_search}, the automated light curve vetting in Section~\ref{sec:vetting}, and the transit fitting analysis in Section~\ref{sec:transit_fitting}. We discuss some final vetting of our candidates, including identifying nearby blend scenarios, in Section~\ref{sec:final_vetting}. Our giant planet candidates are presented in Section~\ref{sec:candidates} and discuss some preliminary spectroscopic follow-up in Section~\ref{sec:espresso}. We discuss the injection-recovery simulations used to determine the detection efficiency of our pipeline in Section~\ref{sec:injrecov_tests} and present our occurrence rate measurements in Section~\ref{sec:occ_rates}, comparing them to previous occurrence rate studies.

\section{TESS Low-Mass Star Sample}\label{sec:lmstar_sample}
\begin{figure}
    \centering
    \includegraphics[width=0.95\columnwidth]{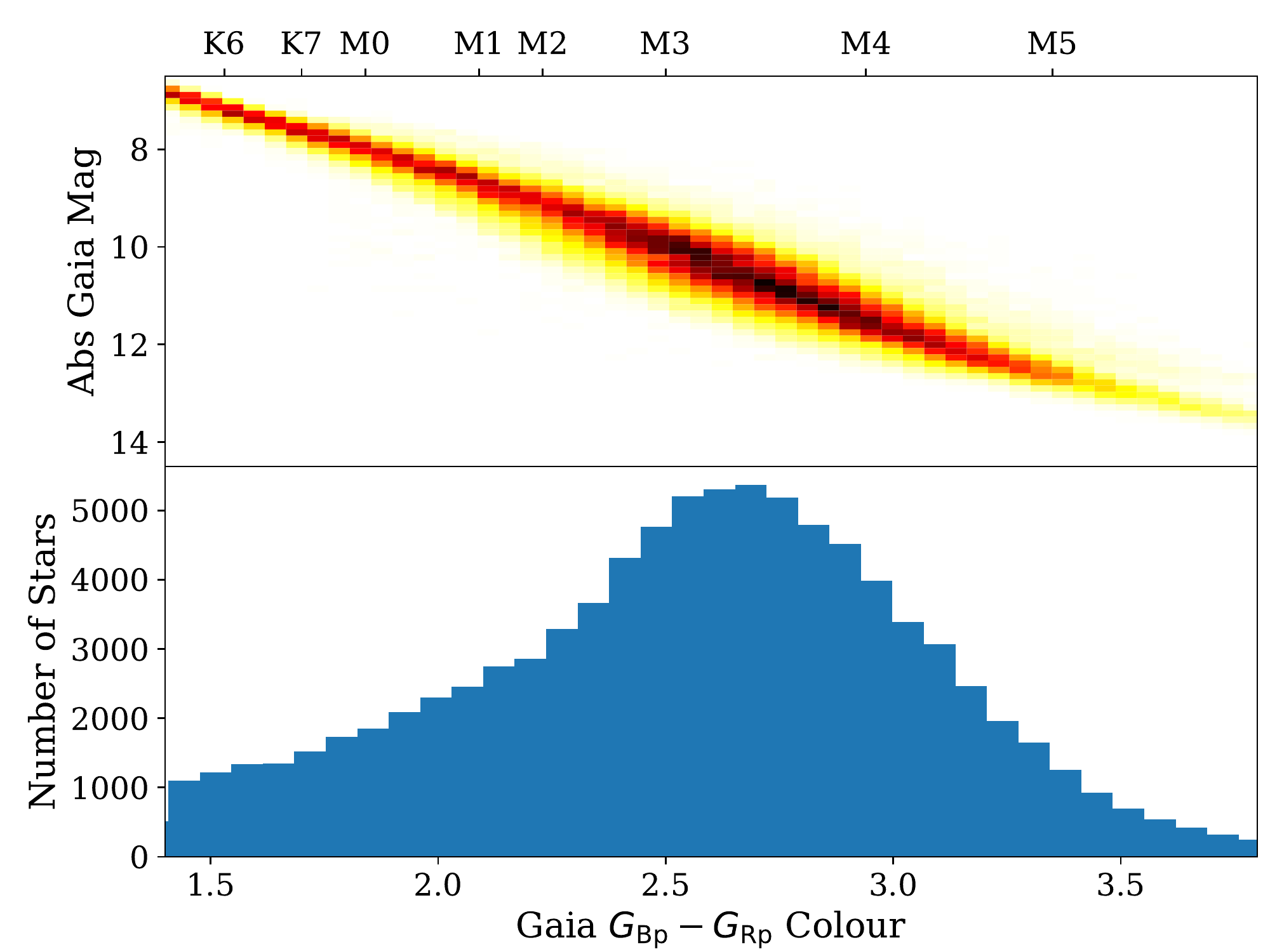}
    \caption[\textit{Gaia} colour-magnitude diagram]{\textbf{Top:} \textit{Gaia} colour-magnitude diagram showing the stellar type distribution of the 91,306 low-mass stars in our sample (density heat map). \textbf{Bottom:} Histogram showing the \textit{Gaia} colour distribution of the sample.}
    \label{fig:lmstar_hr}
\end{figure}
\begin{figure*}
    \centering
    \includegraphics[width=0.9\textwidth]{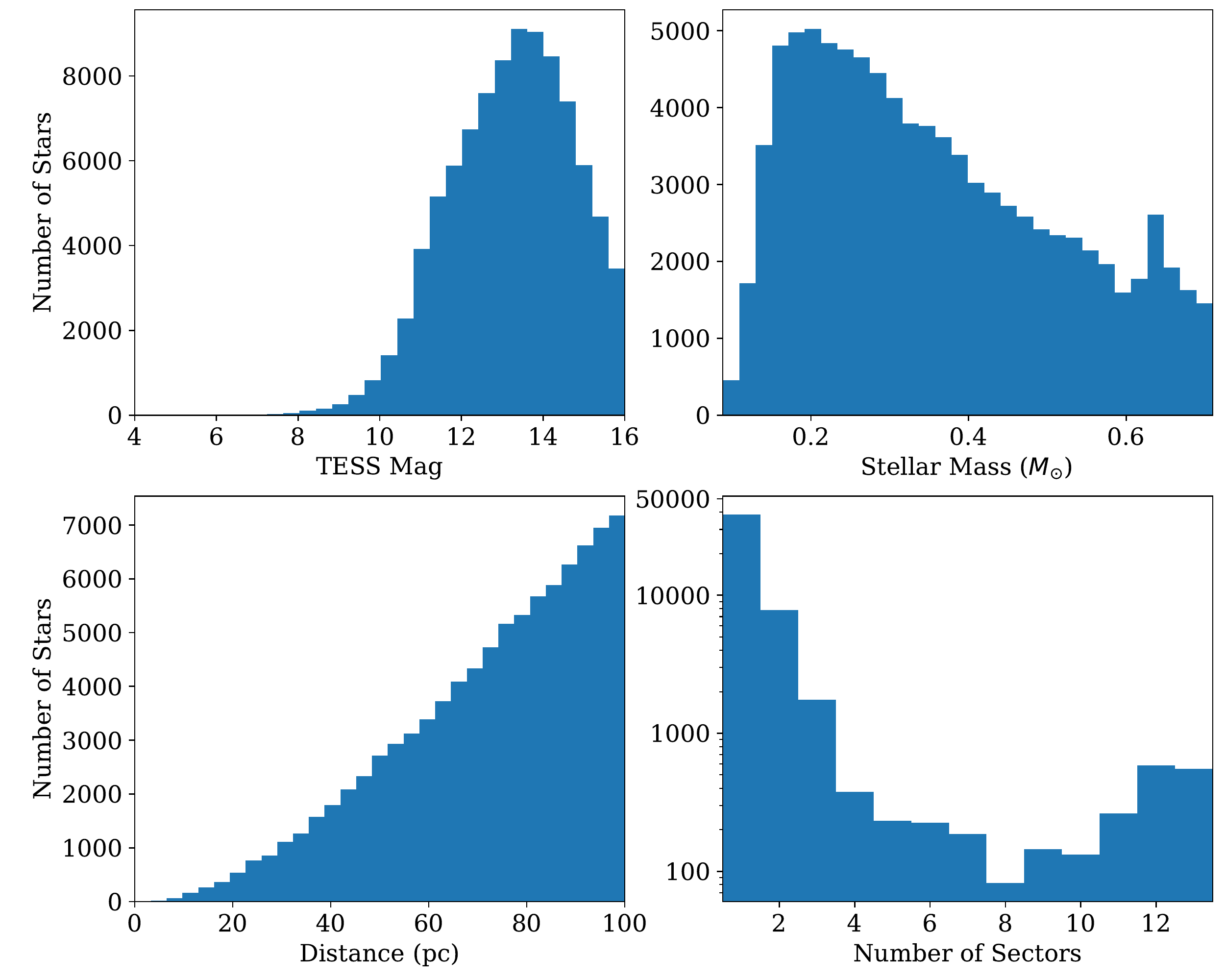}
    \caption[Histograms of TESS low-mass star sample]{Histograms displaying the apparent magnitude (in the TESS bandpass), stellar mass, and distances for the 91,306 stars in our low-mass star sample. We also show the number of TESS sectors for which as TESS-SPOC 30\,minute cadence light curve exists for each object, plotting this histogram on a log scale.}
    \label{fig:sample_hist}
\end{figure*}
Launched in 2018 the Transiting Exoplanet Survey Satellite \citep[TESS;][]{ricker2015tess} has been surveying the whole sky for transiting planets, providing us with a new way to study the planetary occurrence rates for different types of host star. TESS observes the sky in sectors. During each sector a 24\,degree $\times$ 96\,degree area of the sky is observed for 27.4\,days. Across the primary mission the northern and southern hemispheres were observed for thirteen sectors each. These sectors overlapped, especially around the ecliptic poles, and so while most objects receive one or two sectors of observations, objects close to the ecliptic poles can receive up to 13 sectors of coverage.

The primary science goal of the TESS mission is to discover small radius planets -- $\rpl \leq 4$\,\rearth\ -- amenable to further characterisation including with \textit{JWST} \citep{gardner2006jwst}. The discovery of these small exoplanets was primarily achieved through 2\,minute cadence light curves produced for a selected sample of stars from each sector. In addition to these high cadence light curves, the TESS Full-Frame-Images (FFIs) are supplied at a longer cadence. During the primary mission this cadence was 30\,minutes. These FFIs allow for the production of long cadence light curves for large samples of stars across the whole sky. These large datasets allow TESS photometry to be used for occurrence rate studies.

We use the 30\,minute cadence light curves from Years 1 and 2 (sectors 1-26) of the TESS mission produced from the TESS FFIs by the TESS-SPOC team \citep{caldwell2020tessspoc}, which we access from the Mikulski Archive for Space Telescopes (MAST) as a High Level Science Product\footnote{TESS-SPOC light curves accessed as a bulk download from \href{https://archive.stsci.edu/hlsp/tess-spoc}{https://archive.stsci.edu/hlsp/tess-spoc}}. These light curves are selected as they provide a high quality reduction and a light curve data product that is largely free of instrumental systematics. In addition, the target selection criteria used by the TESS-SPOC team prioritises low-mass stars such as those we are targeting in this study. The red-sensitive design of the instrument enables TESS to achieve high precision photometry for M and K spectral type stars. Therefore, in this work we seek to use these TESS data to place the best constraints to date on the occurrence of giant planets orbiting low-mass stars.

To select a statistical sample of low-mass stars, we used the TESS Input Catalog version 8 \citep[\tic;][]{stassun2019ticv8}. The \textit{Gaia} DR2 parameters are incorporated into \tic\ allowing for low-mass stars to be more reliably identified and our sample to be free of giant stars. We select stars for our sample which meet all the following criteria:
\begin{enumerate}
    \item $T_{\rm eff} \leq 4500$\,K,
    \item $R_{*} \leq 0.75\ R_{\odot}$,
    \item $d \leq 100$\,pc, and
    \item $T \leq 16$\,mag,
\end{enumerate} 
where \teff\ is the effective temperature of the star, \rstar\ is the stellar radius, $d$ is the distance of the star from the Sun, and $T$ is the stellar apparent magnitude in the TESS filter. The distance cut is imposed as the TESS-SPOC target selection process uses the same selection cut. We find a total of 168,837 stars in the TIC that satisfy all these criteria, and using \textsc{tess-point} \citep{burke2020tesspoint} we identify that 121,402 of those stars were observed during the TESS Prime Mission.
Cross-matching our full TIC sample with the target lists for the TESS-SPOC 30\,minute cadence FFI light curves yields 91,306 stars with a light curve. We present our low-mass star sample in Figure~\ref{fig:lmstar_hr}. From this plot, as well as the histogram in Figure~\ref{fig:sample_hist}, we see that our low-mass star sample spans a wide range of spectral types and stellar masses ($\approx 0.1 - 0.7$\,\msun), with the majority of a sample being lower-mass than 0.4\,\msun\ and so expected to be incapable of forming giant planets \citep[e.g.][]{burn2021ngpps}. Therefore our stellar sample is ideal for studying the extremes of giant planet formation and investigating the lowest-mass host stars around which it is possible to form giant planets. The stars which were observed by TESS but do not have a TESS-SPOC FFI light curve are preferentially faint ($T \lesssim 14$\,mag) and small ($\rstar\ \lesssim 0.25\,\rsun$) stars. While it is important to be aware of this selection we do not expect this to have a substantial impact on our occurrence rate measurements.

From the histogram in Figure~\ref{fig:lmstar_hr} we see that the majority of the stars in our sample have the spectral type M2 to M4. This distribution arises from the apparent magnitude and distance criteria, as well as the distribution of stars across the sky. The left-hand side of the spectral type distribution follows the mass function for stars and is a result of us using a volume limited sample and later spectral type stars being more common in the Galaxy \citep{chabrierbaraffe2000lowmassstars}. The turnover and decrease in numbers to very late spectral types is a result of the magnitude limit of our sample ($T\leq 16$\,mag) and the very low luminosity of M5 and later spectral types stars. 

The apparent magnitude distribution of our sample peaks around $T = 13$\,mag (see top-left histogram in Figure~\ref{fig:sample_hist}). With a magnitude limited sample, we would expect this distribution to continue to rise until the magnitude limit. The turnover arises as a result of the distance limit imposed on our sample. Early M-dwarf stars with apparent magnitudes of $T \gtrsim 13.5$\,mag have distances $> 100$\,pc. The distribution of the number of observing sectors for each star, shown in the bottom right histogram in Figure~\ref{fig:sample_hist}, arises from the TESS observing strategy. As can be seen from the TESS observations footprint\footnote{See \url{https://tess.mit.edu/tess-year-2-observations/}}, the majority of stars receive one or two sectors of coverage. Moreover, due to the overlapping of sectors in the continuous viewing zones surrounding the Ecliptic poles, a larger fraction of stars receive $\geq 11$~sectors of coverage than receive between 6 and 10 sectors of observations.

\section{Transit Search}\label{sec:transit_search}
Using the TESS-SPOC 30\,minute cadence light curves, we firstly exclude any data points with a quality flag $q > 0$. These flags highlight any observations which are adversely affected by cosmic rays and scattered light, among others. For this anal
ysis we use the PDC\_SAP flux time series, which has been largely corrected for instrumental systematics but for which real stellar variability has been preserved \citep{stumpe2012keplerPDC1, stumpe2014pdc2, smith2012keplerPDC2}. To remove stellar variability prior to the planet search we smooth the light curves using a Savitzky-Golay filter, using the implementation from the \textsc{lightkurve} package \citep{lightkurve2018}. The filter uses a window length of 45 data points, approximately 1\,d for 30\,minute cadence data. This window length is long enough to preserve the transits produced by our target systems, which have durations $\ll 1$\,d. 

We then search through the TESS-SPOC light curves for periodic planet candidate transit signals using the Box-fitting Least Squares algorithm \citep[BLS;][]{kovacs2002bls}. We use the \textsc{astropy} implementation of the algorithm\footnote{Docs can be found at \href{https://docs.astropy.org/en/stable/timeseries/bls.html}{https://docs.astropy.org/en/stable/timeseries/bls.html}}. We searched for signals with periods between 1 and 10\,d and durations between 0.03 and 0.3\,d. We identify objects with a Signal Detection Efficiency \citep[SDE; see][]{kovacs2002bls} $\geq 8$ and a transit signal-to-noise (SNR) of greater than 8 as good transiting planet candidates. In total our search yields the detection of 3930 periodic transit-like signals.

\section{False Positive Identification}\label{sec:vetting}

Along with planet candidates, our BLS detections also included a large number of false positives. These can be either real astrophysical false positive scenarios, such as eclipsing binaries or variable stars, or spurious detections. We performed a number of vetting checks to identify these false positive detections and exclude them from our candidate list. Previous transit searches have used similar criteria to identify false positives \citep[eg.][]{montalto2020diamante}. For each of the checks used to identify false positives we employ a strict quantitative cut-off to distinguish between false positives and planetary candidates. Applying these checks to simulated planet signals (see Section~\ref{sec:injrecov_tests}) we find that each check falsely identifies $< 1$\,\% of the simulated planets as false positives. Therefore we are confident that these checks are not excluding a significant number of real planets from our candidate list. To summarise the results of these checks we identify 1828 clear false positives, leaving us with 2102 remaining candidates. Further details on the checks performed and the number of false positives identified by each check are provided in the following section. Note that in the pipeline these checks are performed sequentially in the order they are presented here, and any objects identified as false positives by one check are not passed to the following checks. So the numbers of identified false positives do not include any false positives identified by previous checks.

\subsection{Secondary Eclipse Events}\label{sec:sec_eclipse}
One of the main astrophysical false positive cases for transiting planet searches are eclipsing binary stars \citep[eg.][]{almenara2009corotfalsepositives, santerne2012keplerfalsepositives}. The main signature of an eclipsing binary is the presence of a secondary eclipse. For a circular orbit this eclipse is present at phase 0.5, however for an eccentric orbit this is not always the case. We search for secondary eclipse events present at orbital phases between 0.2 and 0.8. To identify the secondary eclipse events we use a window of width equal to the duration of the BLS transit model and estimate the depth of any eclipse occurring within the window. The median value of the flux values within the window is taken as the depth of the eclipse. We compare the difference between the median flux level inside the window and the median overall out-of-transit flux baseline to the overall RMS scatter of the light curve, \siglc, to assess the significance of any detected signals. While designed to look for eclipsing binaries, this check also identifies a large number of variable stars. This check has the possibility of identifying real planets as false positives in two scenarios. The first is a planet which is detected by BLS at the wrong orbital period, usually at a multiple of two or one and a half times the orbital period. However, while some real planets may be rejected by this check, when considering our injected transiting planets (see Section~\ref{sec:injrecov_tests}) less than 1\,\% of the simulated planets recovered by BLS are detected at incorrect orbital periods which would trigger this check. As such, we do not expect to have missed many, if any, real planets due to this. Moreover, all these effects are captured by the injection-recovery tests (Section~\ref{sec:injrecov_tests}). The second scenario is where the planet itself has a detectable secondary eclipse when it is occulted by the star. However, these planetary eclipses are typically very shallow. As an example, for a 1\,\rjup\ planet in a 1.5\,d orbit around a 0.4\,\msun\ star, using the equations from \citet{alonso2018sececlipses} we estimate a planetary secondary eclipse depth of just 320\,ppm. This is less than the RMS scatter level for the vast majority of our stellar sample. Therefore, while it is important to be aware that these two scenarios could trigger this check, these would be very rare occurrences and we do not expect them to significantly affect our occurrence rate results.

Any object which displays a flux difference greater than 2.0 times \siglc, occurring between phase values of $\phi_{\rm Sec} = 0.47$ and $\phi_{\rm Sec} = 0.53$ is taken to be an eclipsing binary. For the other phases searched for secondary eclipse signals, a minimum flux difference of 3.5 times \siglc\ is required to identify the object as a false positive. A total of 246 objects are identified as false positives by this check.

\subsection{Odd-Even Depth Differences}\label{subsec:oddeven}
Eclipsing binaries can also result in BLS detections with a difference in depth between the odd and even transit events. This occurs when the BLS algorithm folds the light curve on half the true period of the binary, resulting in the primary and secondary eclipses both falling at phase 0. To check for such cases we compare the depth of the odd transits with the depth of the even. The transit depths were estimated by comparing the median of the lowest 20\% in-transit flux values to the median overall out-of-transit flux baseline. Any objects with an odd-even depth difference greater than 3.5\,\siglc\ are identified as likely eclipsing binaries. Poorly sampled transit events can lead to inaccurate depth estimates, causing real transiting planets to be mis-identified as eclipsing binaries. To avoid this, we only apply this check to objects with at least ten in-transit points for both the odd and even transit events. A total of 118 objects are identified as likely eclipsing binaries by this check. 

\subsection{Sector Depth Differences}\label{sec:sectordepthdiff}
\begin{figure}
    \centering
    \includegraphics[width=\columnwidth]{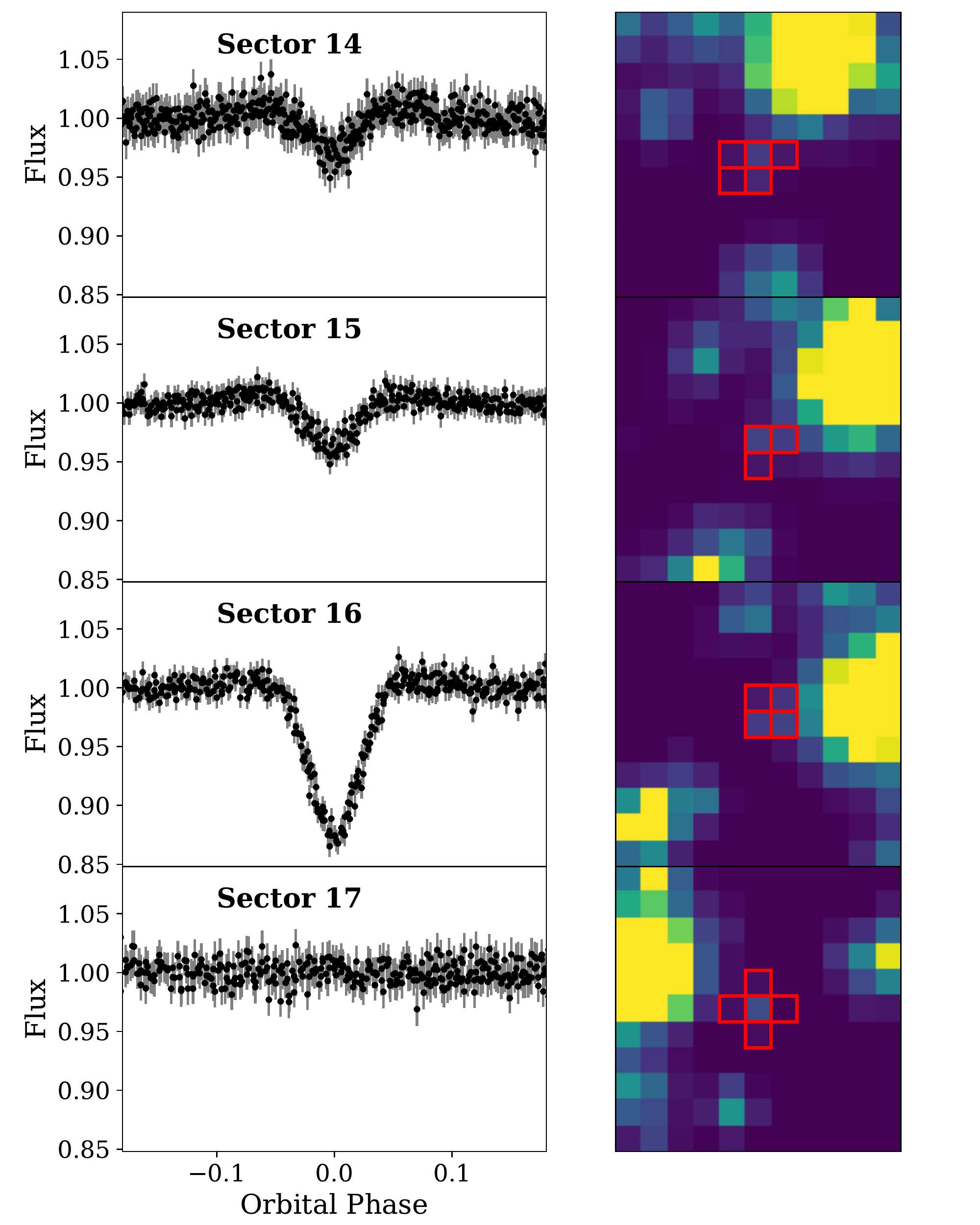}
    \caption[Example of a nearby eclipsing binary identified through a sector depth difference]{\textbf{Left column:} Individual sector light curves for TIC-233684011, which was identified as a nearby eclipsing binary through the difference in eclipse depth observed for different sectors. We plot the spline-smoothed PDC\_SAP light curves phase folded on the best BLS period of $P = 3.64$\,d. \textbf{Right column:} Target pixel cutouts for TIC-233684011 for each sector plotted. The red boxes denote the pixels included in the target aperture used in each sector}
    \label{fig:secdepthdiff}
\end{figure}
In some cases, the depth of the transit events is observed to be different in different sectors, but constant in a given sector. We identified this to be as a result of differing levels of contamination from a nearby eclipsing binary. For different sectors the positions of neighbouring stars relative to a target star will change due to the rotation of the TESS spacecraft (see Figure~\ref{fig:secdepthdiff}).
Moreover, in different sectors a given star will fall in a different location in the camera field-of-view and so will have a differently shaped point-spread-function. The pixel mask used by the TESS-SPOC pipeline to extract photometry also varies from sector to sector. The result of these effects is that the level of contamination from neighbouring stars into the target pixel aperture is different for different sectors. The TESS-SPOC pipeline computes this dilution value and corrects the target light curve for it, so the depths of transits on the target stars should be unaffected by this. However, if one of these neighbouring stars is itself an eclipsing binary then in sectors in which it contributes more light to the target aperture the eclipse event seen in the light curve will be deeper. We present an example of such a scenario in Figure~\ref{fig:secdepthdiff}.

To identify any clear nearby eclipsing binary cases, we compared the depths of the transit events between sectors, for those objects which were observed in more than one sector. The transit depths were calculated using the same method as in Section~\ref{subsec:oddeven}. Any object in which a depth difference between sectors of greater than 3.5\,\siglc\ was identified as a nearby eclipsing binary. A total of 44 objects are identified as likely nearby eclipsing binaries by this method and an example of such an object is shown in Figure~\ref{fig:secdepthdiff}.

\subsection{Transit Phased Variability}\label{sec:phased_variablity}
In addition to secondary eclipses and eclipse depth differences, many eclipsing binaries show out-of-transit variability in phase with the eclipses. This variability can arise from a number of effects, including ellipsoidal modulation of reflection from the an orbiting body \citep{faiglermazeh2011BEER}. While these effects do occur for close orbiting planets, the amplitude of the resulting variability is much larger for eclipsing binaries. 

\begin{figure}
    \centering
    \includegraphics[width=\columnwidth]{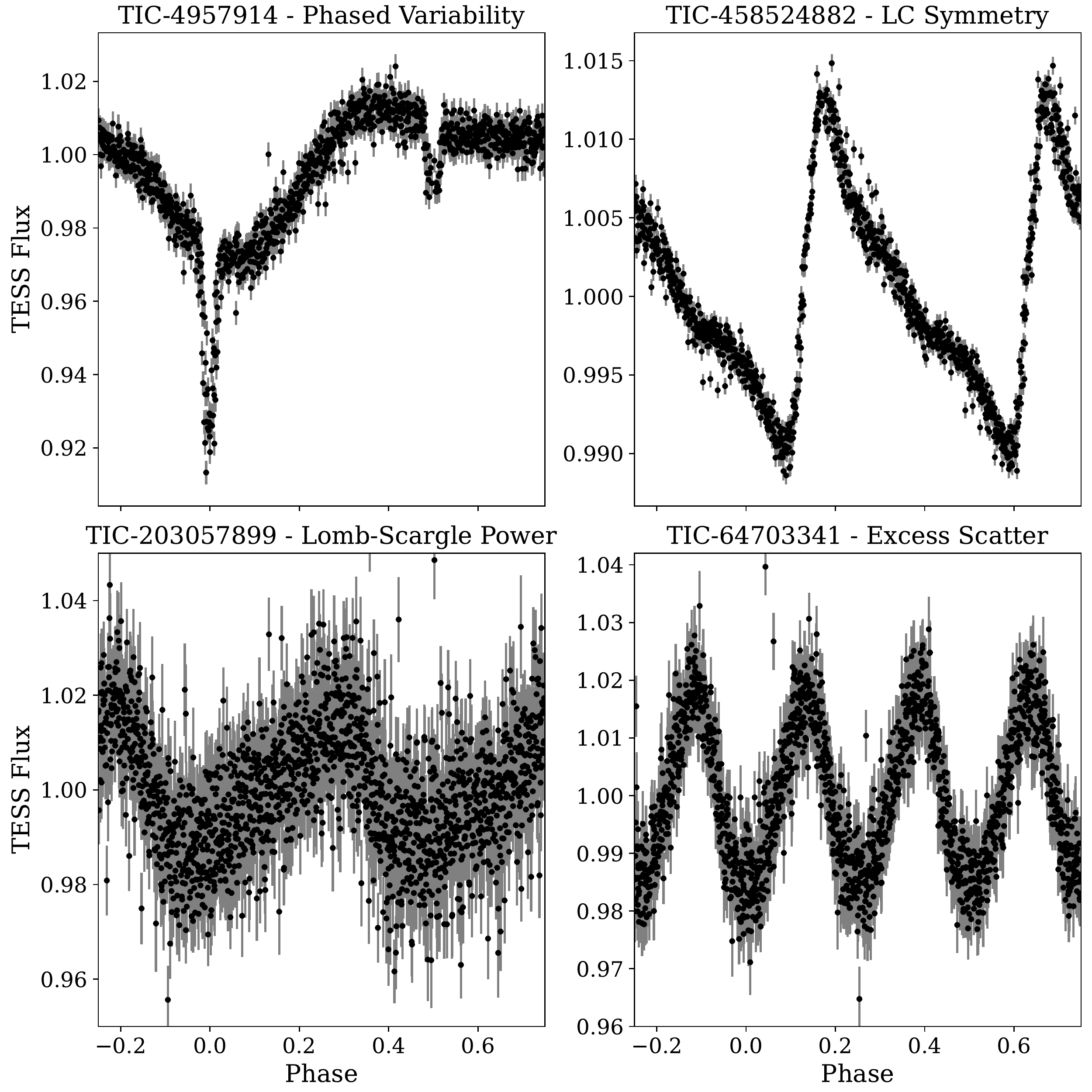}
    \caption[Examples of variable stars identified by the vetting checks]{Examples of false positives identified by the variability analysis checks detailed in Sections~\ref{sec:phased_variablity}~to~\ref{sec:xs_scatter}. The titles of each panel gives the object plotted as well as the vetting check that identified the object as a false positive. Note that for TIC-4957914 the pre-flattened light curve is plotted, but for the rest the light curve smoothing has been applied. All objects are phase folded using the BLS period.}
    \label{fig:variables}
\end{figure}

We search for these variability signals in the pre-flattened light curves as we expect them to be removed by our pre-transit search light curve flattening. We also use the results from the BLS search to mask out the transit events and perform this analysis only on the out-of-transit data. We use the harmonic model presented by \citet{montalto2020diamante}
\begin{equation}
    f_{\rm Harmonic} = A\cos\left(\frac{2\pi\,t}{P}\right) + B\sin\left(\frac{2\pi\,t}{P}\right) + C
\end{equation}
and fit for the coefficients A, B, C for each light curve. We then compare the quality of this harmonic fit to a simple flat line fit by computing the value
\begin{equation}
    R^2_{\rm Harmonic} = 1 - \frac{\Sigma^N_{i=1}\left(f_i - f_{\rm Harmonic, {\it i}}\right)^2}{\Sigma^N_{i=1}\left(f_i - C\right)^2} .
\end{equation}
The better the harmonic fit is to the data, the closer to 1 this value becomes. As with the secondary eclipse check (Section~\ref{sec:sec_eclipse}) this check finds purely variable stars, as well as eclipsing binaries. We identify any object with $R^2_{\rm Harmonic} > 0.5$ as an eclipsing binary or a variable star, and in total identify 231 false positives using this method. An example of an eclipsing binary identified using this check is shown in the top left panel of Figure~\ref{fig:variables}. 

\subsection{Variable Stars}
Besides eclipsing binary stars, the other major astrophysical contaminant in our BLS detections are variable stars. This stellar variability arises from two main mechanisms. The first is the combination of active regions on the stellar surface and the rotation of the star. These active regions are either regions which are less luminous than the majority of the stellar surface, which are called spots, or more luminous, known as faculae. These spots and faculae rotate into and out of view as the star rotates, resulting in a fluctuation in the observed brightness of the star \citep{boisse2012soapI}. Periodic variability in the brightness of stars can also arise from the radial oscillation of the outer layers of the star, known as stellar pulsations \citep{cox1980stellarpulsationtheory}. Multiple types of pulsating stars are found, including $\delta$-Scuti \citep[eg.][]{rodriguez2001deltascuti} and RR-Lyrae \citep[eg.][]{simon1982rrlyrae}, and can exhibit variability with an amplitude on the order of a magnitude. While our low-mass target stars themselves will not exhibit these pulsations, if a nearby star that is contaminating the photometric aperture is such a pulsator then it can imprint such variability onto the target light curve.

\subsubsection{Light curve symmetry}
The phase-folded light curve of an exoplanet transit will be symmetric around phase of zero. Variable star light curves which trigger BLS will often not share this symmetry. As such, by determining the symmetry of the folded light curve we can identify variable stars. We calculate the mean point-to-point scatter of the folded light curve in two configurations. The first is a standard fold running from phase -0.5 to 0.5. The second runs from phase 0.0 to 0.5, where we have taken the absolute value of negative phases. We calculate the ratio of these two values and identify any objects with a ratio greater than 2.5 as variable stars. An example of a variable star identified through the asymmetry of its light curve is shown in the top right panel of Figure~\ref{fig:variables}. A total of 8 objects are identified as variable stars using this analysis.

\subsubsection{Lomb-Scargle Power}
We also check for continuous variability by performing a Generalised Lomb-Scargle \citep{lomb76, scargle82} analysis on the light curve, after masking out the flux data points which are detected as ``in-transit'' by BLS. Any star with a normalised Lomb-Scargle power of greater than 0.3 is identified as a variable star and excluded from our candidate list. An example of such a variable star is shown in the bottom left panel of Figure~\ref{fig:variables}. A total of 1039 objects are identified as variable stars using this test. This check alone removes 26.4\% of the transiting candidates detected by the BLS search. However when applied to the simulated planet light curves just 0.22\% of the simulated planets are excluded by it. Therefore we can be confident that we are not excluding a large number of real planets with this analysis.

\subsubsection{Excess standard deviation scatter metric}\label{sec:xs_scatter}
A further way to identify stellar variability is to compare the RMS scatter of the phase-folded light curve to the root-mean-square of the point-to-point scatter. For a transiting planet light curve, the out-of-transit section of the light curve will be close to flat and dominated by Gaussian noise. Therefore, these two values will differ by a factor of $\sqrt{2}$, with the raw RMS being the smaller of the two. For a variable star, the RMS scatter will be significantly larger than the point-to-point scatter, as the RMS will be dominated by the variability. Therefore, we calculate following metric
\begin{equation}
    \sigma_{\rm excess} = \frac{1}{\sigma_{\rm LC}\sqrt{2}}\sqrt{\frac{\Sigma_{1}^{N} (f_{i} - f_{i-1})^2}{N}}
\end{equation}
where \siglc\ is the RMS scatter and the square-root term gives a measure of the mean point-to-point scatter of the flux. We identify any star with $\sigma_{\rm excess} \leq 0.5$ as a variable star. An example of a variable star identified using this check is shown in the bottom right panel of Figure~\ref{fig:variables}. A total of 12 objects are identified as variable stars through this analysis.

\subsubsection{Depth metric}
As well as astrophysical false positives, our BLS sample also contains a few spurious detections. The most common of these spurious detections are those which are driven by the presence of a small number of outlying points. These outliers can be produced by sharp changes in the scattered light on the camera, or from a decrease in the quality of the spacecraft pointing.
To identify these signals, we calculate the median flux level of the ``in-transit'' data points of the BLS event and compare this value to the reported BLS depth. The BLS depth is determined by a fit to the data, and so is biased to larger values by the outlying points. For real transit events, these two values will be comparable, however for spurious detections being driven by a few outlying points the BLS depth will be significantly larger than the median flux level. We calculate the ratio of the median flux level to the BLS depth and exclude any object with this ratio less than 0.5 as a spurious event. A total of 130 objects are identified as likely spurious events using this method.

\section{Transit Fitting Analysis}\label{sec:transit_fitting}

Having identified and rejected the false positives, we then fit the transit events for the remaining candidates. We use the \textsc{emcee} package \citep{foremanmackey2013emcee} to perform a Markov Chain Monte Carlo (MCMC) analysis. 

The free parameters we use in our analysis are: a reference mid-transit time, $T_{\rm C}$, the planetary orbital period, $P$, the planet-to-star radius ratio, \rpl/\rstar, the scaled semi-major axis, a/\rstar, and the orbital inclination, $i$. We also fit for a free flux baseline offset, $f_0$, which is defined such that the out-of-transit flux is equal to $1 + f_0$. We use a quadratic limb-darkening law and for the limb-darkening coefficients we fit for the $q_{\it 1}, q_{\it 2}$ parameters from the parameterisation of \citet{kipping2013ld}. This parameterisation ensures a physically realistic limb-darkening model for the star. The priors used for each parameter are provided in Table~\ref{tab:fit_priors}. These priors are selected to ensure that we fit physically realistic transit models to each light curve but also that we do not bias the results in any other way.

\begin{table}
    \centering
    \begin{tabular}{|l|c|}
    \hline
    Parameter     & Prior  \\
    \hline
     & \\
    Transit centre time, $T_{\rm C}$  &  $\mathcal{U}(T_{\rm C, \ BLS} - 0.1\,P_{\rm BLS};$ \\
    & $T_{\rm C, \ BLS} + 0.1\,P_{\rm BLS})$ \\
     & \\
    Orbital Period, $P$    &  $\mathcal{U}\left(0.95\,P_{\rm BLS}; 1.05\,P_{\rm BLS}\right)$ \\
     & \\
    Planet-to-star radius ratio, \rpl/\rstar & $\mathcal{U}\left(0; 1\right)$ \\
     & \\
    Scaled semi-major axis, a/\rstar & $\mathcal{U}\left(1.1; \infty\right)$ \\
     & \\
    Orbital inclination, $i$ & $\mathcal{U}\left(0^{\circ}; 90^{\circ}\right)$ \\
     & \\
    Limb-darkening parameter, $q_{\it 1}$ & $\mathcal{U}\left(0; 1\right)$ \\
     & \\
    Limb-darkening parameter, $q_{\it 2}$ & $\mathcal{U}\left(0; 1\right)$ \\
     & \\
    Constant flux offset, $f_0$ & $\mathcal{U}\left(-0.05; 0.05\right)$ \\
     & \\
    \hline
    \end{tabular}
    \caption[Parameter priors used in the transit fitting]{Priors used for each free parameter in the transit fitting detailed in Section~\ref{sec:transit_fitting}. Note $\mathcal{U}\left(A; B\right)$ denotes a uniform prior with lower bound A and upper bound B.
    }
    \label{tab:fit_priors}
\end{table}

For each object, 24 independent chains were each sampled for 7,500 steps, following a burn-in phase of 2,500 steps. This resulted in posterior distributions of 180,000 samples for each candidate. We use the results from our transit MCMC analysis to assess the likely planetary nature of our candidates.

As a first step we consider the set of parameters for each object which resulted in the highest log likelihood values during the fitting; referred to in this paper as the ``best-fit'' parameters. We include only objects with a best-fit planetary radius in the range $0.6\,\rjup \leq \rpl \leq 2.0\,\rjup$ in our candidate list. Using the transit models computed using the best-fit parameters we also calculate the ratio $q_{\rm trans} = T_{dur} / P$, where $T_{dur}$ is the transit duration. We accept only objects with \mbox{$0.0 < q_{\rm trans} \leq 0.1$} as likely planetary candidates. We test the threshold chosen for this criterion by calculating the $q_{\rm trans}$ value for all the planets we simulate for the Injection-Recovery tests (see Section~\ref{sec:injrecov_tests}). The highest $q_{\rm trans}$ value is 0.079, with 90\% of the simulated planets having $q_{\rm trans} < 0.03$. Therefore, our chosen threshold is such that real transiting giant planets are unlikely to be excluded by this criterion. Using these two parameter threshold criteria, we identify just 93 of the 2102 candidates as being strong giant planet candidates.

For a final automated step we use the posterior distributions from the MCMC sampling to assess the likely planetary nature of our candidates. We determine a posterior distribution for \rpl\ using the \rpl/\rstar\ posterior and the host star radius from the TIC. Similarly, we use the posteriors for $P$ and a/\rstar\ along with Kepler's 3$^{\rm rd}$ law to determine a posterior distribution for the stellar density, \rhostar. We note that these \rhostar\ values are calculated assuming a circular orbit. With these distributions in \rpl\ and \rhostar\ we then determine what fraction of the posterior falls in the giant planet regime, which we define with the following parameter ranges: $0.6\,\rjup \leq \rpl \leq 2.0\,\rjup$ and 1.5\,g\,cm$^{-3} \leq \rhostar \leq 200$\,g\,cm$^{-3}$. We then exclude from our candidate list any object with this fraction $f_{\rm GP} < 15$\%. From these automated steps outlined above, we identified a list of 44 candidate giant exoplanets. 

\section{Planet candidate manual vetting}\label{sec:final_vetting}
We now perform a number of manual checks for each of the candidates found by the automated search in order to ensure we have a final list of strong candidates. 

\subsection{Visual Inspection}
For the first manual check we visually inspected the TESS-SPOC 30\,minute cadence light curves for each of our 44 candidates. From this visual inspection, we identified 22 of our candidates which displayed clear signs of being astrophysical false positives or due to systematic effects. Examples of this included secondary eclipses and odd-even transit depth differences which had amplitudes below the automated thresholds and sharp ramps in flux at the end of TESS orbits which were folded into data gaps to mimic transit signals. These ramps are the result of increased scattered light at the end of the TESS orbit being imperfectly corrected for. These 22 systems are then removed from our candidate list.

\subsection{Independently Confirmed Objects}\label{sec:known}
Three of our candidates have had their natures confirmed prior to this work, which we independently detected with our pipeline. Two of these are the giant planets TOI-1130\,c \citep[TIC-254113311; 0.974\,\mjup;][]{huang2020toi1130} and WASP-107\,b \citep[TIC-429302040; 0.12\,\mjup;][]{anderson2017wasp107}. The third object is the known brown dwarf LP~261-75\,b \citep[TIC-67646988; 68\,\mjup;][]{irwin2018toi1779}. Therefore we exclude TIC-67646988 from our candidate list.

\subsection{Blend Scenario Checks}\label{sec:blends}
Due to the large pixel scale of TESS (21\arcsec/pix), we must ensure that the transit signals we observe are not due to a nearby star blending into the target aperture. Using the Target Pixel Files (TPFs) from the TESS-SPOC pipeline it is possible to inspect each candidate to search for clear signs of a blend scenario. We download the TPFs for each candidate using the \textsc{lightkurve} Python software \citep{lightkurve2018}. We then perform two tests to search for evidence of the source of the signal being a neighbouring star and not the target.

The first of these involves generating light curves using different aperture sizes. We compare three different aperture sizes. The first is the aperture used by the TESS-SPOC pipeline. The second is a small aperture of just the single pixel at the location of the target star, and the third is a large $5\times 5$ pixel aperture also centred at the target star location. By comparing the depths of these three light curves, we can reveal blend scenarios. These blend scenarios will in general be cases where a larger aperture results in a deeper transit, signifying that the signal is arising from a nearby star that is only partially within the target aperture, but more fully within the larger aperture. If all three light curves have equal depths then it is unlikely the signal arises from a nearby star. Similarly, a shallower transit for the larger aperture also points towards the signal arising from the target star. This is because the result of increasing the aperture size here is simply the inclusion of an increased amount of dilution, and so the signal being on any of the nearby stars is unlikely. The second test we performed is to generate light curves for each individual pixel in a $7 \times 7$ pixel grid centred on the target star location. With these light curves we can investigate whether the transit signals are clustered around the target star, or are offset.

From these tests, one of our candidates - TIC-174440134 - displayed clear signs of being a nearby eclipsing binary. The depth of the transit events was clearly correlated with the size of the photometric aperture used, and the individual pixel light curves showed a clear offset of the centre of the event from the target star. This object was therefore removed from our candidate list.

\subsection{TESS higher cadence light curves}\label{sec:tess2min}
For some of our remaining candidates, we have access to higher cadence TESS photometry, which can often provide more information on the nature of a candidate. 

For five of our candidates this higher cadence photometry is in the form of a TESS 2\,minute cadence SPOC light curve, from which we determined the following about these candidates. The 2\,minute cadence light curve for TIC-156067195 revealed a subtle difference in both the depth and the phase position of the odd and even transit events, identifying this candidate as an eclipsing stellar binary system with a slightly eccentric orbit. For TIC-190885165 the 2\,minute cadence photometry reveals more clearly the shape of the transit events. By refitting this photometry we find a best fit radius of 2.59\,\rjup\ for the companion revealing its nature as most likely a stellar companion. TIC-389900760 and TIC-38460940 were initially considered as candidates because from the 30\,minute cadence alone they could have been giant planets on grazing orbits. The 2\,minute cadence photometry shows that they are not grazing, and so are most likely real planets but with $\rpl < 0.6\,\rjup$, and so smaller than the giant planets we are searching for in this work. This is consistent with them being identified as TOI-2120.01 ($\rpl = 2.58\pm0.24$\,\rearth) and TOI-805.01 ($\rpl = 1.28\pm0.62$\,\rearth). Based on this we exclude them both from our giant planet candidate list. For TIC-178709444 the 2\,minute cadence light curve contains no evidence that the candidate is a false positive. In fact the higher cadence confirms the transit as non-grazing ($b < 1$) providing us with increased confidence that the transit signal is produced by a planet and is not the result of an eclipsing binary.

For eight more of our candidates - TIC-46432937, TIC-67512645, TIC-95112238, TIC-165227846, TIC-202468443, TIC-243641947, TIC-271321097, and TIC-335590096 - there are TESS-SPOC 10\,minute cadence full-frame-image light curves from the extended mission available. Using these data we can identify TIC-271321097 as an eclipsing binary system. The extended mission light curve clearly displays primary and secondary eclipses. The secondary eclipses are what is seen in the primary mission 30\,minute cadence light curve, with the primary eclipses falling into the data gaps. TIC-271321097 is therefore excluded from our candidate list. For TIC-67512645 the 10\,minute cadence photometry revealed the transit as being flat-bottomed, whereas in the 30\,minute cadence photometry there was the possibility of a grazing eclipse, thereby strengthening the likelihood of this candidate being a real planet. Similarly for TIC-243641947 and TIC-335590096 the 10\,minute cadence photometry confirmed the transits as flat-bottomed, further supporting the nature of these two objects as planetary candidates. The 10\,minute cadence photometry for the remaining four objects contains no evidence that they are false positives, and so they remain as good transiting giant planet system candidates. 

\begin{figure}
    \centering
    \includegraphics[width=\columnwidth]{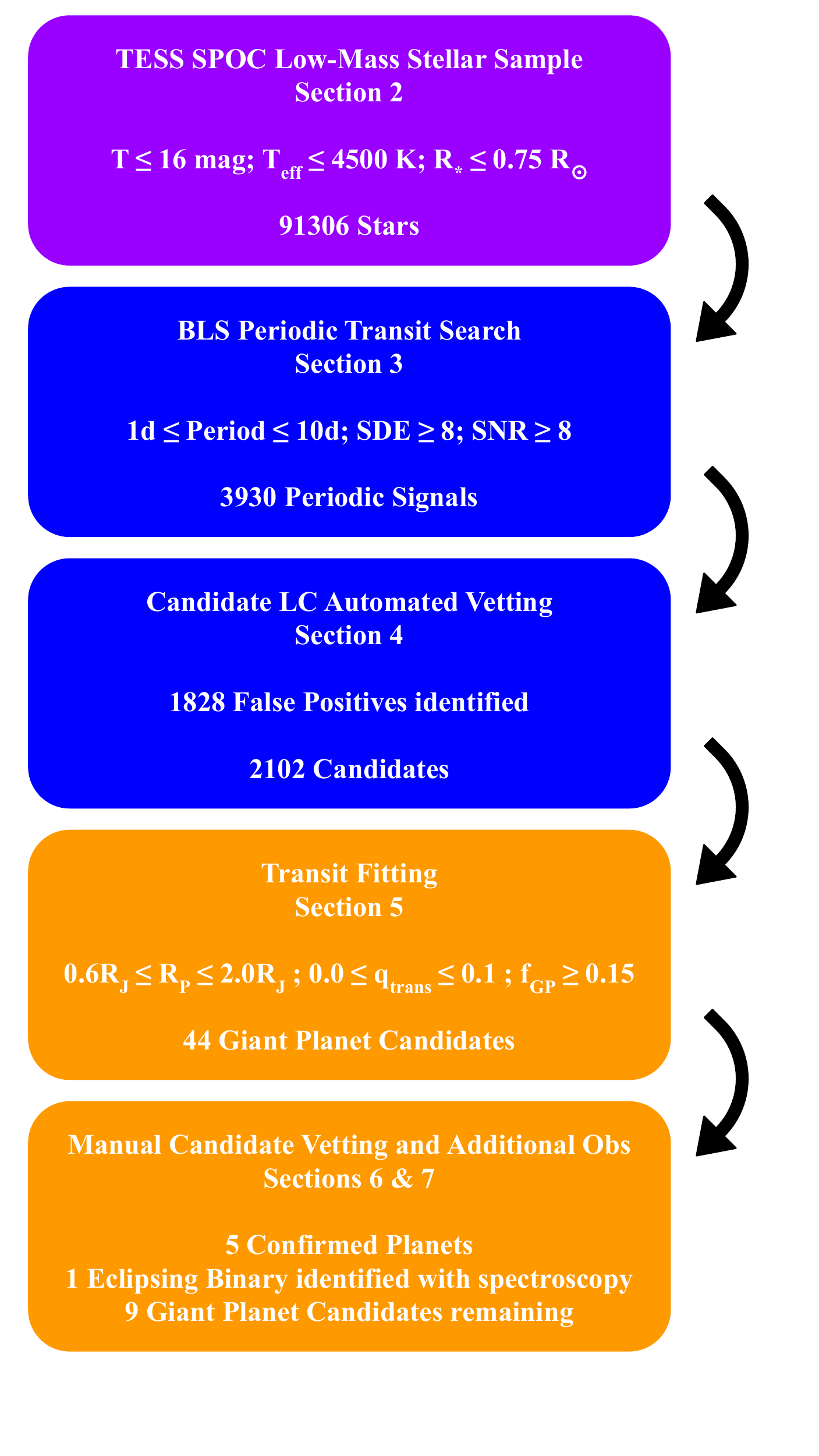}
    \caption[Flow chart of the steps of the planet search pipeline]{Flow chart of the steps of the planet search pipeline. The numbers indicate the number of stars from our sample for each step.}
    \label{fig:flow_chart}
\end{figure}

\section{Giant Planet Candidates}\label{sec:candidates}
\begin{figure*}
    \centering
    \includegraphics[width=\textwidth]{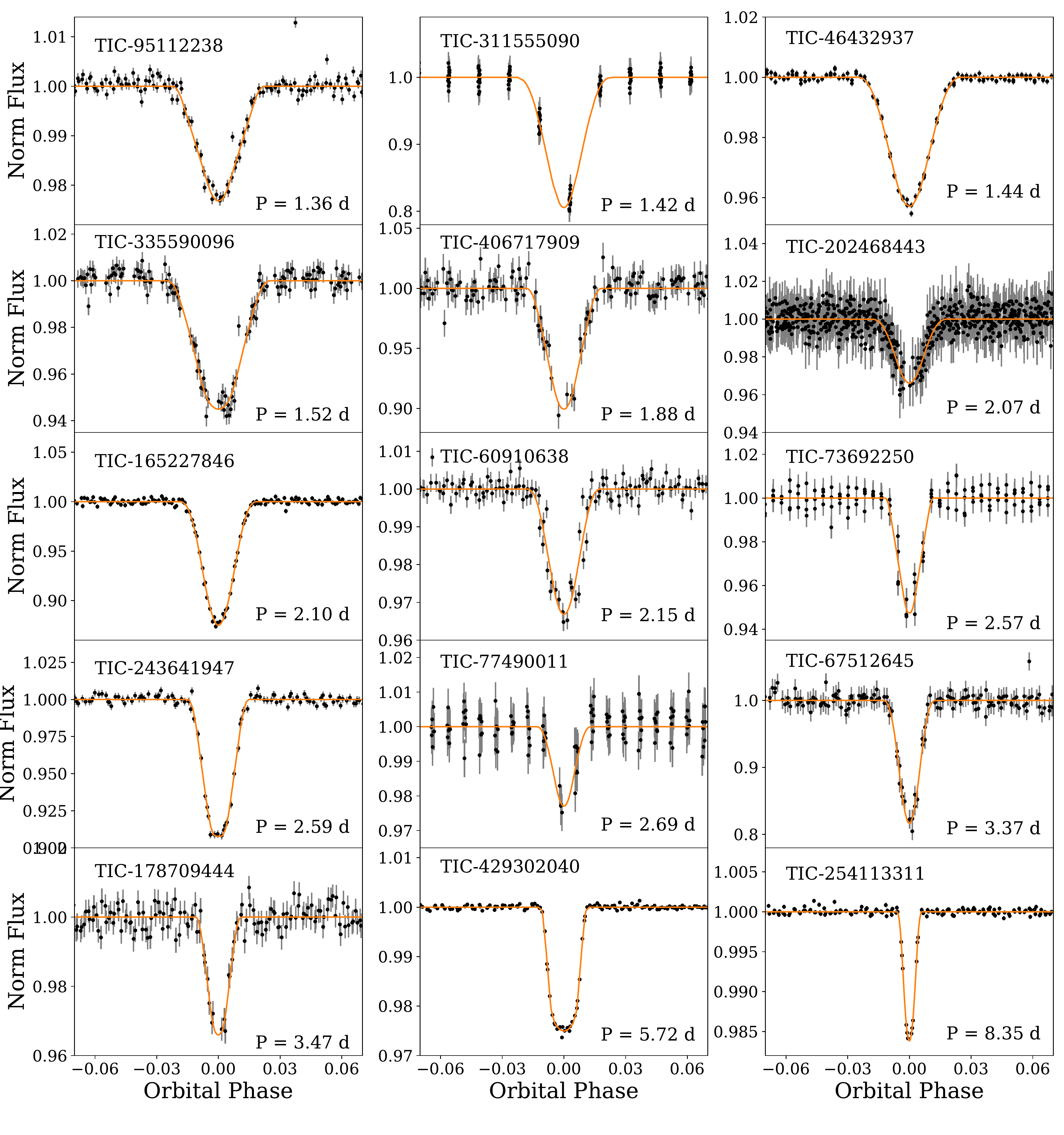}
    \caption[Giant planet candidate transit light curves]{Light curves for the fifteen transiting giant planet candidates phase folded using the ephemerides from the fitting detailed in Section~\ref{sec:transit_fitting} and zoomed in to show the transit features. The orange lines show the best fit models. The TIC IDs and orbital periods are given for each system.}
    \label{fig:cand_lcs}
\end{figure*}
\begin{figure}
    \centering
    \includegraphics[width=0.95\columnwidth]{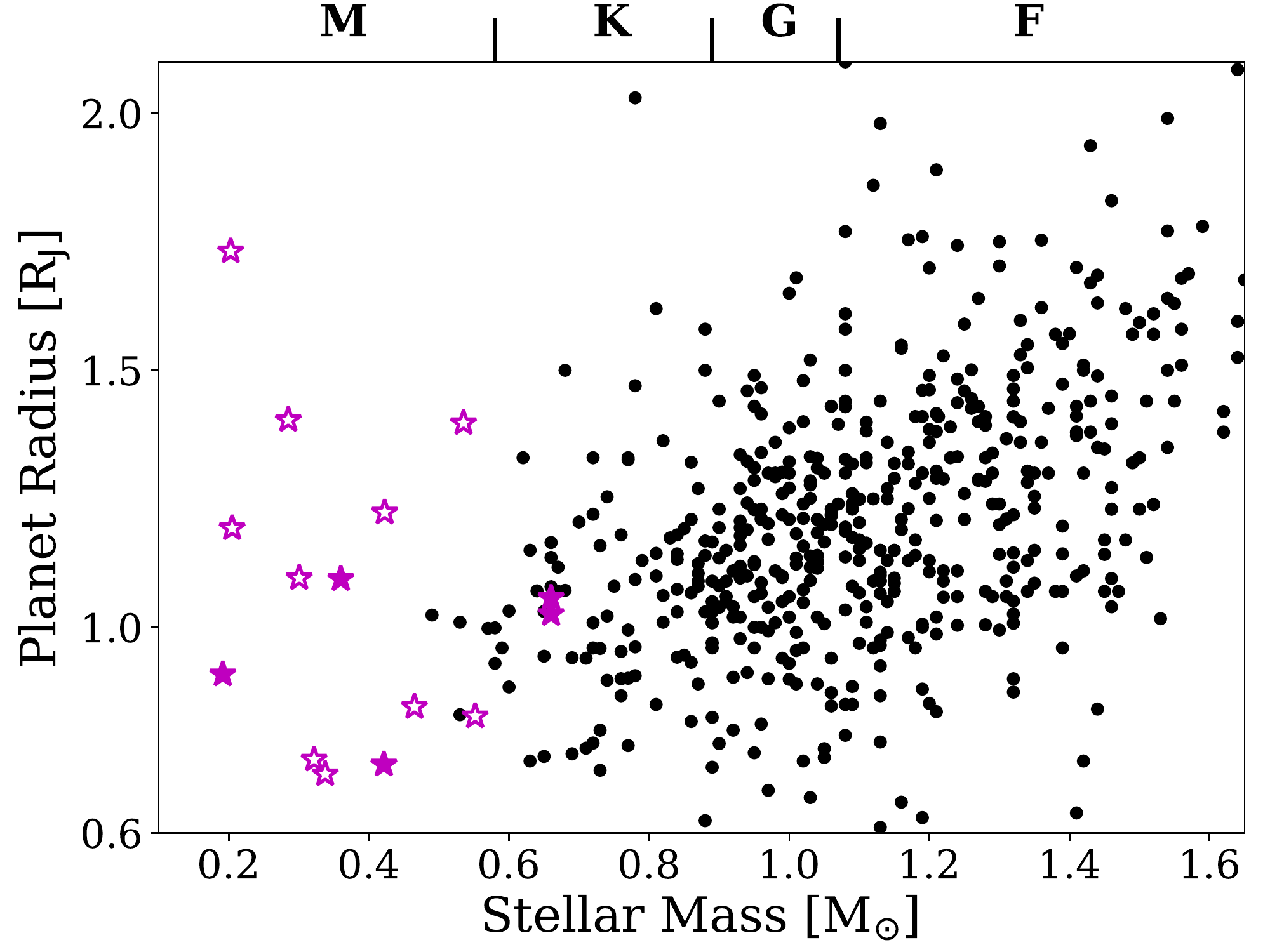}
    \caption[Giant planet candidates compared to known transiting gas giant exoplanets]{Radii of our giant planet candidates compared to the known transiting exoplanets plotted in Figure~\ref{fig:ms_mpl} as a function of the mass of the host star. The magenta stars show the fifteen gas giant planet candidates detected in this study, with the filled stars denoting the five confirmed giant planets.  For these candidates the \rpl\ values plotted are the best-fit values from the fitting detailed in Section~\ref{sec:transit_fitting}}
    \label{fig:ms_mpl_wcands}
\end{figure}
\begin{figure}
    \centering
    \includegraphics[width=0.95\columnwidth]{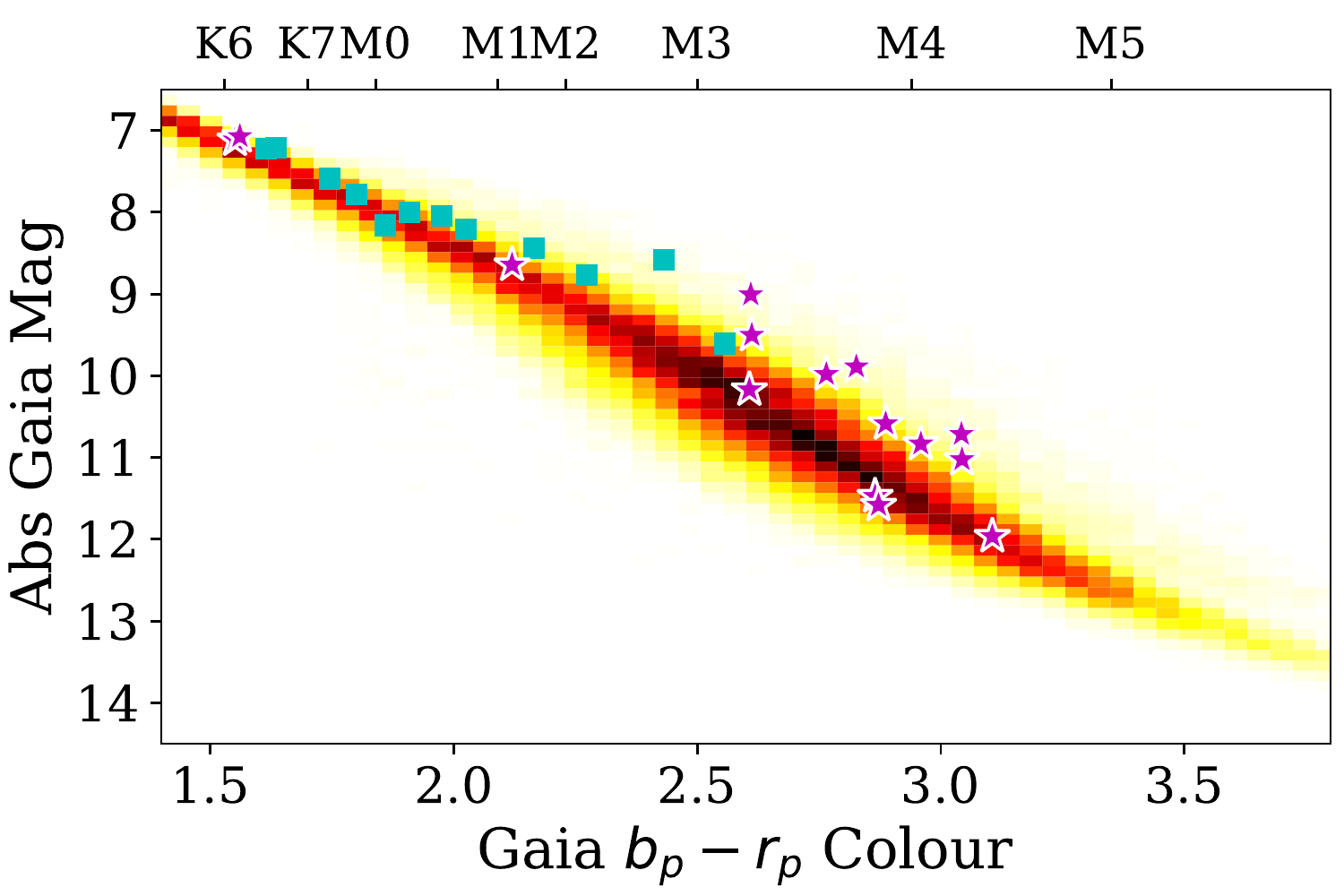}
    \caption[Comparison of giant planet candidates to the low-mass star \textit{Gaia} colour-magnitude distribution]{Comparison of our fifteen giant planet candidates (magenta stars) to the \textit{Gaia} colour-magnitude distribution of the 91,306 low-mass stars in our sample (density heat map). We also plot the twelve known transiting gas giant planets with hosts stars with masses $\leq 0.65\,\msun$ as the blue squares for comparison to our candidates.}
    \label{fig:lmstar_hr_wcands}
\end{figure}
From our full planet search we have identified a final selection of fifteen giant planet candidates. The numbers of objects from our sample at each step of the planet search pipeline are summarised in Figure~\ref{fig:flow_chart}.  The planetary parameters derived for our candidates in Section~\ref{sec:transit_fitting} are provided in Table~\ref{tab:planet_params} and we display the transit events for each candidate in Figure~\ref{fig:cand_lcs}. We provide details on the host stars of our candidates in Tables~\ref{tab:stellar_mags}~and~\ref{tab:stellar_params}. Two of our giant planet candidates are already confirmed as giant planets and four have been identified as TOIs, for which we provide details in Table~\ref{tab:followup}. A further two candidates have been independently reported as CTOIs; these are TIC-95112238 \citep{montalto2020diamante} and TIC-77490011 \citep{montalto2023diamante2}. The remaining seven of our giant planet candidates are new candidates that were not previously known as planets or identified as candidates prior to our search.

We also compare our sample of candidates to the full population of known transiting giant planets in Figures~\ref{fig:ms_mpl_wcands} and to the subset of known transiting giant planets with low-mass host stars in \ref{fig:lmstar_hr_wcands}. From these two figures, it is evident that our search has extended the population of known transiting gas planets to lower stellar mass hosts than ever before. From Figure~\ref{fig:ms_mpl_wcands} we can see that our giant planet candidates have host stars spanning almost the full stellar mass range of our sample. We have detected eight candidates whose host stars have masses $\mstar < 0.4\,\msun$, with some as low-mass as $\mstar \approx 0.2\,\msun$. If confirmed as real giant planets, these candidates will provide strong tension with our current understanding of planet formation \citep[e.g.][]{pascucci2016ppdiskmasses, burn2021ngpps}.

From the colour-magnitude diagram in Figure~\ref{fig:lmstar_hr_wcands}, we can see that some of our targets appear at slightly brighter absolute magnitudes than the bulk population, for a given Gaia colour. Such an effect could be indicative of these objects being binary stars, with the additional star resulting in the objects appearing much brighter than would be expected for a single star of the same spectral type. However, we also note that one of the known giant planets, HATS-74A b \citep{jordan2022hats74_77}, also sits raised above the line. This system consists of a low-mass star that hosts a transiting giant planet, as well as a bound stellar companion on a wide orbit. Such wide stellar companions have been found to be common for hot Jupiter host stars \citep{knutson2014hjfriends1, ngo2016hjfriends4}. As such, we cannot exclude these objects as giant planet candidates simply based on their position on the colour-magnitude diagram.

The \textit{Gaia} astrometric re-normalised unit weight error \citep[RUWE; ][]{lindegrenGaiaRUWE} can also be used to indicate the potential presence of a stellar binary, with a values of ${\rm RUWE} > 1.4$ often used to signify the presence of a stellar companion in the system \citep{lindegren2021ruweDR3A, lindegren2021DR3RUWEB}. For our candidates we identify two with ${\rm RUWE} > 1.4$. The first of these is TIC-429302040 with ${\rm RUWE} = 1.54$, which is the known exoplanet system WASP-107 \citep{anderson2017wasp107}. There is a known outer planetary companion in the system \citep[WASP-107\,c; 1088\,d; $\mpl\,\sin i = 0.36\,\mjup$; ][]{piaulet2021wasp107c} although there is no known stellar companion in the system. The second is TIC-311555090 with ${\rm RUWE} = 2.29$. Similarly with above, this object cannot be excluded as a giant planet candidate based solely on this RUWE value, but we report it here for reference.

\subsection{ESPRESSO Radial Velocity Monitoring}\label{sec:espresso}
We note that not all of candidates may be true giant planets. In order to uncover their true natures, we have begun a program to obtain mass measurements for our candidates using the ESPRESSO spectrograph \citep{pepe2020espresso} on the VLT. The current status of the spectroscopic follow-up of our candidates is summarised in Table~\ref{tab:followup}. ESPRESSO has been successfully used over recent years to confirm similar systems \citep[eg. HATS-71 b; ][]{bakos2020hats71}.  We have been reducing the ESPRESSO data throughout the program using the publicly available pipeline, which runs in the ESOReflex environment \citep{esoreflex2013}. 

The full spectroscopic follow-up for all of our candidates is beyond the scope of this work, and is likely a couple of years at least away, however here we report some results to date that are consequential for this study. To date we have obtained spectroscopic observations for four of our candidates: TIC-243641947 (TOI-3235), TIC-178709444 (TOI-762), TIC-67512645, and TIC-60910638. 
For TIC-243641947 we have a completed spectroscopic orbit confirming the transiting companion as a giant planet. The full analysis of this system is outside the scope of this paper and is the subject of a recent paper \citep{hobson2023toi3235}. For TIC-178709444 and TIC-67512645 the spectroscopic data we have obtained to date also shows the transiting companions to be giant planets, although a small amount of further spectroscopic observations are required to fully confirm both. For the occurrence rate analysis in this paper, we therefore take these three candidates to be genuine giant planets. For TIC-60910638 the spectroscopic observations reveal its true nature to be an eclipsing binary system and therefore exclude it from our occurrence rate analysis, although we include it in our reported candidate list for completion.

\begin{table*}
    \centering
        \begin{tabular}{| m{3.05cm} | m{4cm} m{4cm} m{4cm} |}
    \hline
    & & & \\
 & \textbf{TIC-95112238} & \textbf{TIC-311555090} & \textbf{TIC-46432937} \\ 
 & & & \\
\tc\ (BJD - 2457000) & $1468.539538\pm0.000541$ & $1817.344136\pm0.001122$ & $1468.630246\pm0.000325$ \\
  Period (d) & $1.362066\pm0.000060$ & $1.416762\pm0.000112$ & $1.440455\pm0.000037$ \\
\rpl\ (\rjup) & $0.79^{0.03}_{-0.04}$ & $1.33^{0.61}_{-0.34}$ & $1.57^{1.23}_{-0.45}$ \\
\rprs & $0.14^{0.01}_{-0.01}$ & $0.58^{0.27}_{-0.15}$ & $0.30^{0.23}_{-0.09}$ \\
\rhostar\ (\gccc) & $17.03^{4.61}_{-5.24}$ & $30.54^{8.95}_{-6.54}$ & $5.45^{1.38}_{-0.46}$ \\
a/\rstar & $11.9^{1.0}_{-1.4}$ & $14.8^{1.3}_{-1.1}$ & $8.4^{0.7}_{-0.2}$ \\
$i$ (deg) & $88.1^{1.4}_{-1.5}$ & $87.0^{1.8}_{-1.1}$ & $83.5^{2.1}_{-1.8}$ \\

    & & & \\
 & \textbf{TIC-335590096} & \textbf{TIC-406717909} & \textbf{TIC-202468443} \\ 
 & & & \\
\tc\ (BJD - 2457000) & $1571.796562\pm0.000817$ & $1684.513652\pm0.001532$ & $1712.155769\pm0.002109$ \\
 Period (d) & $1.522755\pm0.000095$ & $1.877180\pm0.000214$ & $2.068584\pm0.000027$ \\
\rpl\ (\rjup) & $0.76^{0.14}_{-0.04}$ & $1.55^{1.38}_{-0.33}$ & $1.38^{1.09}_{-0.80}$ \\
\rprs & $0.22^{0.04}_{-0.01}$ & $0.37^{0.33}_{-0.08}$ & $0.46^{0.37}_{-0.27}$ \\
\rhostar\ (\gccc) & $8.37^{2.82}_{-4.36}$ & $13.64^{6.59}_{-3.55}$ & $6.11^{4.12}_{-2.03}$ \\
a/\rstar & $10.1^{1.0}_{-2.2}$ & $13.6^{1.9}_{-1.3}$ & $11.1^{2.1}_{-1.4}$ \\
$i$ (deg) & $88.0^{1.4}_{-3.7}$ & $86.6^{2.5}_{-1.8}$ & $83.6^{2.8}_{-1.8}$ \\

    & & & \\
 & \textbf{TIC-165227846} & \textbf{TIC-60910638} & \textbf{TIC-73692250} \\ 
 & & & \\
\tc\ (BJD - 2457000) & $1572.424716\pm0.000288$ & $1764.926115\pm0.001158$ & $1791.488380\pm0.001541$ \\
 Period (d) & $2.096696\pm0.000041$ & $2.149809\pm0.000176$ & $2.572953\pm0.000300$ \\
\rpl\ (\rjup) & $1.14^{0.60}_{-0.10}$ & $0.94^{1.29}_{-0.17}$ & $0.76^{1.40}_{-0.06}$ \\
\rprs & $0.36^{0.19}_{-0.03}$ & $0.21^{0.28}_{-0.04}$ & $0.23^{0.42}_{-0.02}$ \\
\rhostar\ (\gccc) & $12.13^{1.86}_{-1.21}$ & $5.67^{10.35}_{-1.49}$ & $21.11^{19.55}_{-8.58}$ \\
a/\rstar & $14.1^{0.7}_{-0.5}$ & $11.1^{4.6}_{-1.1}$ & $19.5^{4.8}_{-3.1}$ \\
$i$ (deg) & $87.4^{0.8}_{-1.4}$ & $85.6^{3.0}_{-2.5}$ & $88.2^{1.3}_{-2.6}$ \\

    & & & \\
 & \textbf{TIC-243641947} & \textbf{TIC-77490011} & \textbf{TIC-67512645} \\ 
 & & & \\
\tc\ (BJD - 2457000) & $1603.695100\pm0.000444$ & $1791.395003\pm0.004345$ & $1901.057603\pm0.001274$ \\
 Period (d) & $2.592465\pm0.000109$ & $2.687991\pm0.000772$ & $3.370806\pm0.000296$ \\
\rpl\ (\rjup) & $1.07^{0.03}_{-0.03}$ & $0.86^{0.98}_{-0.52}$ & $1.33^{0.57}_{-0.43}$ \\
\rprs & $0.29^{0.01}_{-0.01}$ & $0.38^{0.43}_{-0.23}$ & $0.61^{0.26}_{-0.20}$ \\
\rhostar\ (\gccc) & $16.55^{2.01}_{-2.90}$ & $17.15^{78.12}_{-11.84}$ & $20.47^{5.18}_{-4.02}$ \\
a/\rstar & $18.0^{0.7}_{-1.1}$ & $18.7^{14.4}_{-6.1}$ & $23.1^{1.8}_{-1.6}$ \\
$i$ (deg) & $89.3^{0.5}_{-0.7}$ & $86.7^{2.5}_{-2.8}$ & $87.7^{1.1}_{-0.6}$ \\

    & & & \\
 & \textbf{TIC-178709444} & \textbf{TIC-429302040} & \textbf{TIC-254113311} \\ 
 & & & \\
\tc\ (BJD - 2457000) & $1572.678689\pm0.001707$ & $1574.147012\pm0.000308$ & $1657.903824\pm0.000449$ \\
 Period (d) & $3.471712\pm0.000474$ & $5.721615\pm0.000157$ & $8.351069\pm0.000350$ \\
\rpl\ (\rjup) & $1.51^{1.83}_{-0.79}$ & $1.04^{0.02}_{-0.02}$ & $1.95^{2.59}_{-0.89}$ \\
\rprs & $0.36^{0.44}_{-0.19}$ & $0.15^{0.00}_{-0.00}$ & $0.27^{0.36}_{-0.12}$ \\
\rhostar\ (\gccc) & $7.57^{10.32}_{-2.14}$ & $3.31^{0.37}_{-0.45}$ & $2.96^{0.75}_{-0.39}$ \\
a/\rstar & $16.9^{5.6}_{-1.8}$ & $17.9^{0.6}_{-0.9}$ & $22.2^{1.7}_{-1.0}$ \\
$i$ (deg) & $86.1^{2.9}_{-1.4}$ & $89.0^{0.6}_{-0.5}$ & $87.1^{0.8}_{-0.9}$ \\

     & & & \\
     \hline
    \end{tabular}
    \caption[Planetary parameters for our giant planet candidates]{Planetary parameters for our giant planet candidates derived from the transit fitting in Section~\ref{sec:transit_fitting}. The quoted values are the 50$^{\rm th}$ percentiles of the posterior distributions, and the uncertainties are defined by the 16$^{\rm th}$ and 84$^{\rm th}$ percentiles and represent the 1$\sigma$ uncertainty.}
    \label{tab:planet_params}
\end{table*}

\begin{table}
    \centering
    \begin{tabular}{ c c c }
    \hline
       \textbf{TIC} &  \textbf{Radial Velocites} & \textbf{Comments} \\
       \hline
       \noalign{\smallskip}\noalign{\smallskip}
        95112238 &  NO & CTOI \citep{montalto2020diamante} \\
         & & \textsc{triceratops} FPP = 0.24 \\
        \noalign{\smallskip}
        311555090 &  NO & \textsc{triceratops} FPP = 0.62 \\
        \noalign{\smallskip}
        46432937 & NO & \textsc{triceratops} FPP = 0.025 \\
        \noalign{\smallskip}
        335590096 & NO & TOI-4860 \\
         & & \textsc{triceratops} FPP = 1.00 \\
        \noalign{\smallskip}
        406717909 & NO & \textsc{triceratops} FPP = 0.67 \\
        \noalign{\smallskip}
        202468443 & NO & TOI-5268\\
         & & \textsc{triceratops} FPP = 0.94 \\
        \noalign{\smallskip}
        165227846 & NO & \textsc{triceratops} FPP = 1.00 \\
        \noalign{\smallskip}
        60910638 & ESPRESSO & ESPRESSO observations show this\\
        & &  to be an eclipsing binary  \\
        \noalign{\smallskip}
        73692250 &  NO & \textsc{triceratops} FPP = 0.35 \\
        \noalign{\smallskip}
        243641947 &  ESPRESSO & TOI-3235 \\
        & & Radial velocity observations confirm as \\
        & & a giant planet \citep{hobson2023toi3235} \\
        \noalign{\smallskip}
        77490011 &  NO & CTOI \citep{montalto2023diamante2} \\
         & & \textsc{triceratops} FPP = 0.22 \\
        \noalign{\smallskip}
        67512645 &  ESPRESSO & Radial velocity observations to date \\
        & & consistent with a giant planet  \\
        \noalign{\smallskip}
        \multirow[t]{3}{*}{178709444} & ESPRESSO & TOI-762 \\
         & & Radial velocity observations to date \\
        & & consistent with a giant planet  \\
        \noalign{\smallskip}
        429302040 &  \textit{Euler}/CORALIE & WASP-107 \citep{anderson2017wasp107} \\
        \noalign{\smallskip}
        254113311 & CHIRON & TOI-1130 \citep{huang2020toi1130} \\
    \hline
    \end{tabular}
    \caption{We provide a summary of the spectroscopic follow-up data available to date for our giant planet candidates, as well as providing any comments on their natures for the solved systems or the \textsc{triceratops} FFPs for the currently un-dispositioned candidates.}
    \label{tab:followup}
\end{table}

\section{Injection and Recovery Tests}\label{sec:injrecov_tests}
In order to derive occurrence rates for giant planets we must determine the detection efficiency of our search pipeline. To do this we performed injection-recovery simulations, which have been used successfully for studies into \kepler\ results \citep[e.g.][]{christiansen2020keplerinjectiontests}. 

In order to accurately replicate the noise properties and observing windows for the light curves we used real TESS-SPOC 30\,minute light curves, choosing as our base light curve set the selection of our input sample which did not yield a significant BLS detection. 
We then injected the transiting planet signals, simulating 10 different planets for each star. The parameters for each simulated planet were randomly selected from the uniform distributions. The parameters and corresponding distributions are as follows: orbital period, $P$ - selected from a range of $1 - 10$\,d, transit centre time, $T_{\rm C}$ - selected such that the first transit lies with one orbital period of the start of the light curve, planet radius, \rpl\ - selected from a range of $0.6 - 2.0$\,\rjup, and impact parameter, $b$ - selected from a range of $0 - 1 + R_{\rm P} / R_{\ast}$.

The stellar parameters, as obtained from the TIC, were then used along with the randomly drawn parameters to derive the required parameters to generate the \textsc{batman} light curve model. The stellar radius, \rstar, and \rpl\ were used to derive the radius ratio, \rpl\ / \rstar. The stellar mass, \mstar, along with \rstar\ and $P$ were used to derive the scaled semi-major axis of the orbit, $a$ / \rstar using Kepler's Third Law, assuming circular orbits. Finally, $a$ / \rstar\ and $b$ were used to calculate the orbital inclination, $i$. Note that we assume circular orbits for all simulated planets.

Using this method, we simulated 873760 transiting planets, which we then passed through our planet search pipeline, as if they were real light curves. Every simulated light curve was searched using BLS, and the simulated systems which yielded a significant detection at a period within 5\% of the injected period were then passed to the vetting steps we outline in Section~\ref{sec:vetting}. Our pipeline yielded 780379 significant BLS detections from our simulated planet population, of which a total of 759632 simulated planets then successfully passed the vetting checks described in Section~\ref{sec:vetting}. The transit fitting (Section~\ref{sec:transit_fitting}) is a key step in the planet search pipeline, however we lack the computing resources to perform a transit fit on all 759632 simulated planets. Instead, we selected at random a single simulated planet for each star and performed a transit fit for these. By doing this, we were able to obtain a representative estimate of the effect had by the transit fitting on the overall detection efficiency. We do not perform any visual inspection or further manual vetting (see Section~\ref{sec:final_vetting}) on our simulated planets. Therefore the false identification of any real planets as false positives due to this manual vetting is not captured in these detection efficiencies. Given the characteristically strong signals of transiting hot Jupiters compared with the photometric noise and any systematics we do not expect this to significantly affect our results.

\begin{figure*}
    \centering
    \includegraphics[width=\textwidth]{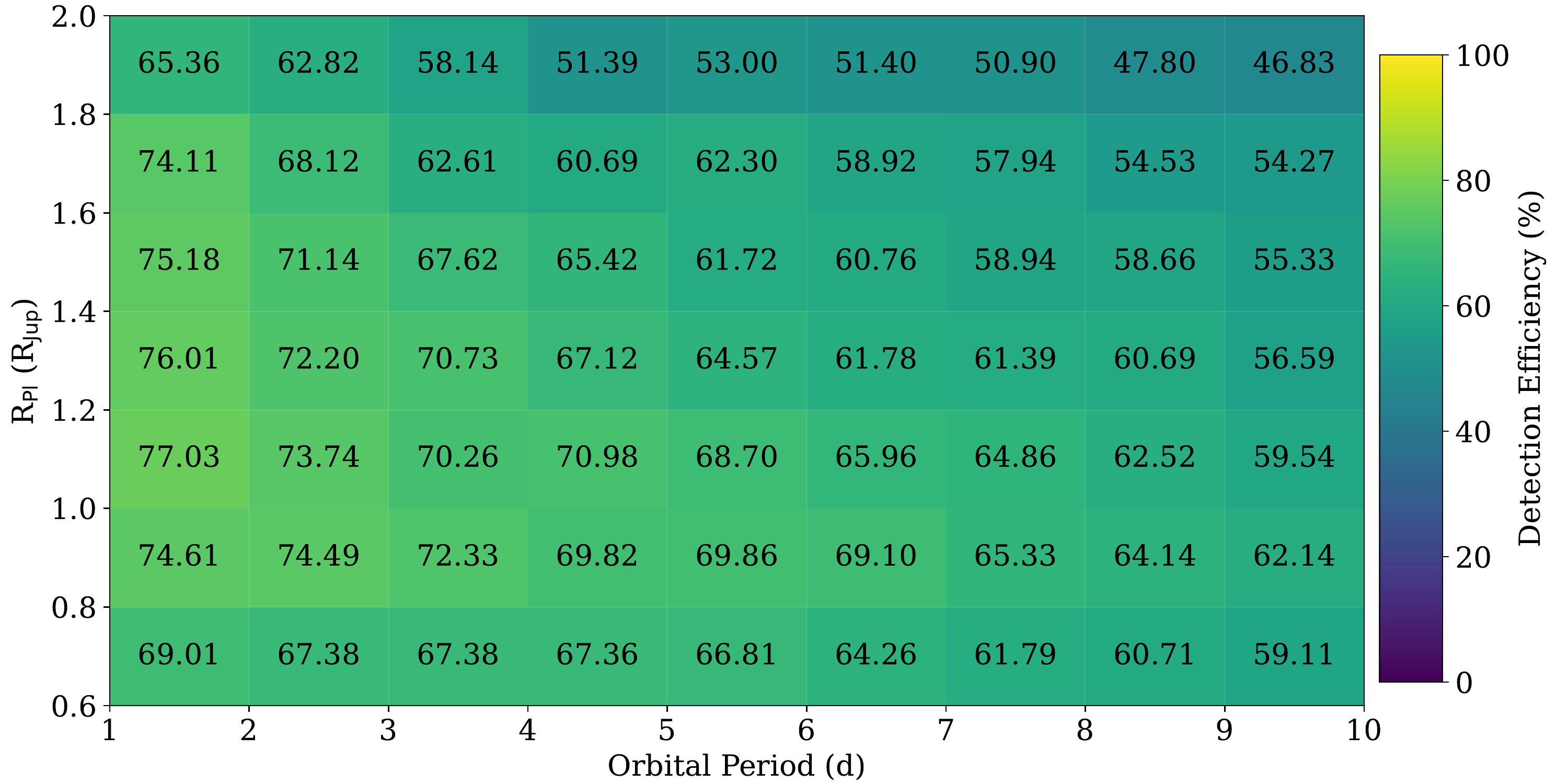}
    \caption{Two-dimensional map showing the detection efficiency of our planet search pipeline and how this varies for different planet radii and orbital periods. These sensitivities take into account the transit search, light curve vetting, and transit fitting analysis steps of the pipeline. The colour of each cell gives the detection efficiency withing that cell. The numbers printed onto each cell also give these detection efficiencies in percent. The values here highlight the sensitivity of our pipeline to transiting giant planets.}
    \label{fig:detect_eff}
\end{figure*}

The detection efficiencies we calculate for our planet search pipeline are presented in Figure~\ref{fig:detect_eff}. We can see that across the majority of the $ \{P; R_{\rm P}\} $ parameter space we study in this search we have high detection efficiencies, and that our sensitivity to transiting planets decreases for longer orbital periods. This is to be expected, as the signal-to-noise ratio of the phase-folded transit event decreases for these planets due to the smaller number of total transit events in the light curve. 

In the large majority of cases where a simulated planet would be excluded from our giant planet candidate list by the transit fitting analysis this is due to the fitting algorithms finding an inaccurate impact parameter, $b$, for these planets, leading to non-grazing transits being fit with grazing transit models and vice-versa. The 30\,minute cadence of the TESS FFI observations reduces our knowledge of the true transit shape causing these inaccurate fits, resulting in either an over- or under-estimation of the planet radius. This has more of an impact on simulated planets with longer orbital periods, as they exhibit fewer transit events, which combined with the 30\,minute cadence of the TESS FFI observations results in a sparse sampling of the transit event in the phase-folded light curve. This sparse sampling then amplifies the lack of knowledge available for the true transit shape, and therefore the true planet radius. Similarly, simulated planets with radii close to our giant planet radius limits of 0.6\,\rjup\ and 2.0\,\rjup\ are more susceptible to be misidentified as false positives due to this miscalculation of the impact parameter, as a smaller inaccuracy in the recovered impact parameter, and thus planet radius, is needed to result in best fit parameters for these simulated planets outside the giant planet regime.

\begin{figure*}
    \centering
    \includegraphics[width=0.9\textwidth]{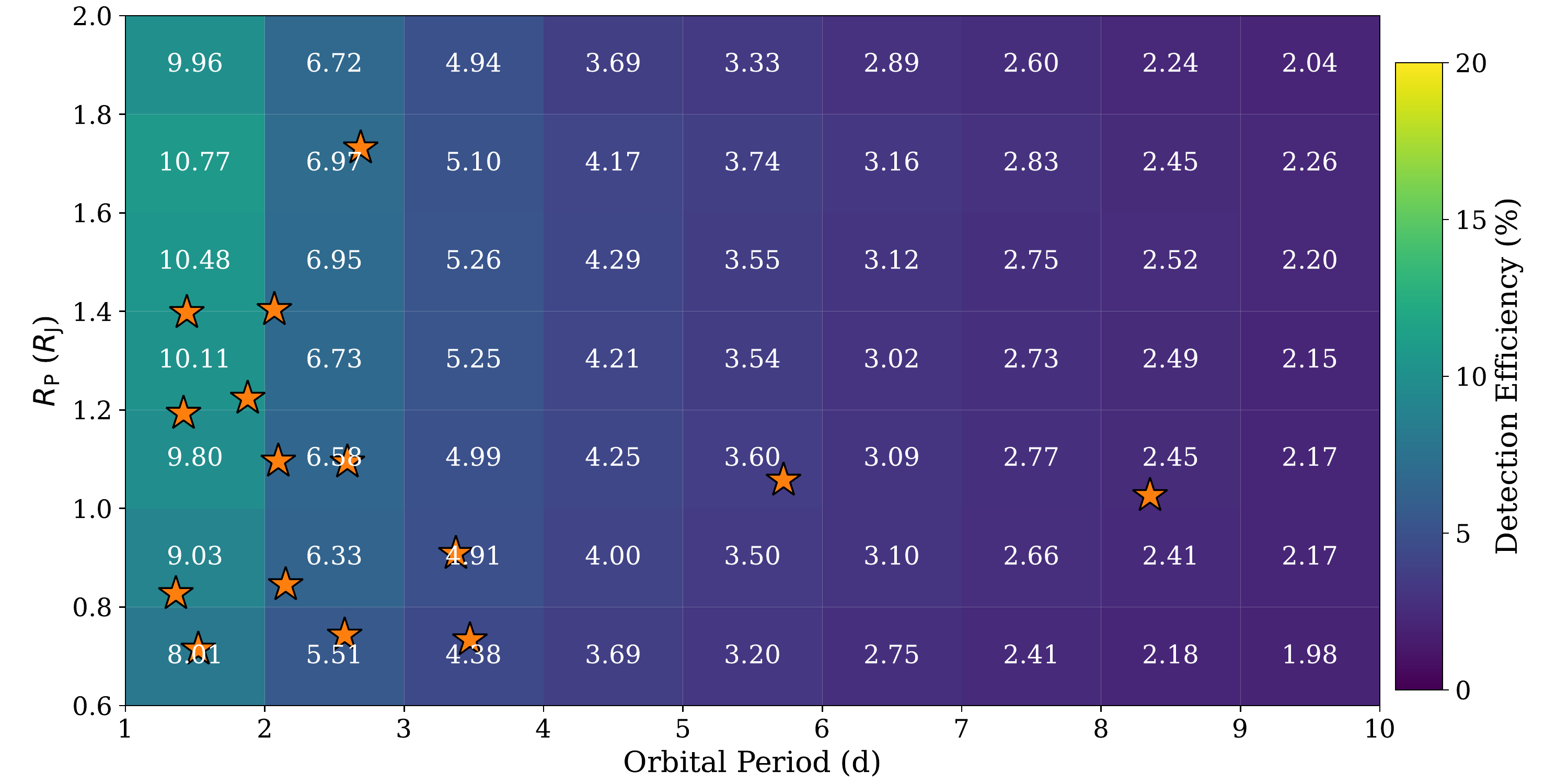}
    \caption{Two-dimensional colour map displaying the overall detection efficiency of the planet search pipeline, split across multiple bins for the injected orbital period and planet radius of the simulated planet. These detection efficiencies take into account all of the pipeline steps as well as the geometric transit probability. The numbers in each bin give the detection efficiency for the corresponding bin in percent. The stars highlight the locations in $\{P; R_{\rm P}\}$ parameter space of our fifteen giant planet candidates.}
    \label{fig:detect_eff_full}
\end{figure*}

From the overall detection efficiency of our pipeline, which is plotted in Figure~\ref{fig:detect_eff_full}, it is clear that there is a strong dependence with orbital period. This arises as the geometric transit probability itself depends strongly on the orbital separation between the star and planet. The detection efficiency of the BLS search and transit fitting analysis also depends on the orbital period of the planet. It is therefore not surprising that the majority of our giant planet candidates are found with orbital periods $P \leq 3.5$\,d.


\section{Giant Planet Occurrence Rates}\label{sec:occ_rates}
\begin{table}
    \centering
    \begin{tabular}{ c c c }
    \hline
    \textbf{Stellar Mass Range (\msun)} & \textbf{Occurrence Rate (\%)} \\
    \hline
    $0.088 - 0.71$ & \Nfoccfullnopc \\
    $0.088 - 0.71$ (Max) & \Nfoccfullmaxnopc \\
    $0.088 - 0.71$ (Min) & \Nfoccfullminnopc \\
    & & \\
    $0.088 - 0.4$ & \Nfoccbelowburnnopc \\
    $0.088 - 0.4$ (Max)& \Nfoccbelowburnmaxnopc \\
    $0.088 - 0.4$ (Min) & \Nfoccbelowburnminnopc \\
    & & \\
    $0.088 - 0.26$ & \Nfocclowmassnopc \\
    $0.26 - 0.42$ & \Nfoccmedmassnopc \\
    $0.42 - 0.71$ & \Nfocchighmassnopc \\
    \hline
    \end{tabular}
    \caption{Summary of the short period giant planet occurrence rates we measure in this work, for different stellar mass ranges. The ``Max'' labels denote the occurrence rates that have been calculated assuming all nine of the currently unconfirmed candidates are real, and the ``Min'' labels denote the occurrence rates that have been calculated assuming none of these nine candidates are real.}
    \label{tab:occrates}
\end{table}
Using the candidates from our planet search and results from our planet injection-recovery simulations, we can now place constraints on the occurrence rates of giant planets orbiting low-mass stars. We can calculate an occurrence rate, \focc, using t
he following equation
\begin{equation}\label{eq:focc}
    f_{\rm occ} = \frac{n_{\rm pl}}{N_{\rm pr}} ,
\end{equation}
where $n_{\rm pl}$ is the number of planets detected and $N_{\rm pr}$ is the number of stars amenable to the discovery of an exoplanet \citep[see e.g.][]{bayliss2011hjfrequency, gan2022tesshjoccurrence}. This factor is calculated as follows
\begin{equation}
    N_{\rm pr} = N_{\ast} \frac{\Sigma_{i=1}^{N_{\rm sim}} \delta_{{\rm vet}, i} \mathcal{P}_{{\rm tr}, i}}{N_{\rm sim}} \frac{\Sigma_{j=1}^{N_{\rm fit}} \delta_{{\rm fit}, j}}{N_{\rm fit}} ,
\end{equation}
where $N_{\ast}$ is the number of stars in the stellar sample, $N_{\rm sim}$ is the total number of planets simulated, $N_{\rm fit}$ is the number of simulated planets which were passed to the transit fitting, $\delta_{{\rm vet}, i}$ is a detection delta function that equals one if the simulated planet was detected by BLS and passed the vetting checks, else equals zero, $\delta_{{\rm fit}, j}$ is a similar delta function dependent on whether a simulated planet was identified as a giant planet candidate by the transit fitting, and $\mathcal{P}_{{\rm tr}, i}$ is the geometric transit probability of the simulated planet. This transit probability is defined as
\begin{equation}
    \mathcal{P}_{{\rm tr}, i} = \frac{R_{\rm P} + R_{\ast}}{a} .
\end{equation}

It is clear from Figure~\ref{fig:detect_eff_full} that both the detection efficiency and planet occurrence rates will be functions of the orbital period, $P$, and planet radius \rpl. Therefore, to calculate the overall giant planet occurrence rate we first calculate individual $f_{\rm occ}$ values for each of the cells in $\{P; R_{\rm P}\}$ parameter space in Figure~\ref{fig:detect_eff_full}, and then calculate the sum of these individual values. Similarly to estimate the uncertainty on the occurrence rate, $\sigma_{f_{\rm occ}}$, for each cell in $\{P; R_{\rm P}\}$ parameter space in which at least one planet candidate is detected we calculate the uncertainties on $n_{\rm pl}$ and $N_{\rm pr}$ as $\sqrt{n_{\rm pl}}$ and $\sqrt{N_{\rm pr}}$ respectively, as estimated from Poisson counting statistics. The individual cell occurrence rate uncertainty is then computed for each cell as
\begin{equation}
    \left(\frac{\sigma_{f_{\rm occ}}\left(P; R_{\rm P}\right)}{f_{\rm occ}\left(P; R_{\rm P}\right)}\right)^2 = \left(\frac{\sigma_{n_{\rm rpl}}\left(P; R_{\rm P}\right)}{n_{\rm rpl}\left(P; R_{\rm P}\right)}\right)^2 + \left(\frac{\sigma_{N_{\rm pr}}\left(P; R_{\rm P}\right)}{N_{\rm pr}\left(P; R_{\rm P}\right)}\right)^2 .
\end{equation}
Due to the large number of simulated planets injected in this work the first term dominates the occurrence rate uncertainty. The individual values for each cell are then summed in quadrature to calculate the overall occurrence rate uncertainty.

For each $\{P; R_{\rm P}\}$ bin, the value used for $n_{\rm pl}$ is calculated as
\begin{equation}
    n_{\rm pl}\left(P; R_{\rm P}\right) = \Sigma_{i=1}^{N_{\rm cands}} 1 - {\rm FPP}_i
\end{equation}
where $N_{\rm cands}$ is the number of candidates in the parameter space range and FPP is a False Positive Probability that we assign to each candidate. The two previously known planets -- TIC-429302040 and TIC-254113311 -- as well as the three planets we have spectroscopically confirmed -- TIC-243641947, TIC--178709444, and TIC-67512645 -- are all assigned ${\rm FPP} = 0$. TIC-60910638, which our spectroscopic follow-up reveals is an eclipsing binary, is assigned ${\rm FPP} = 1$. For each of the remaining candidates, we calculate an FPP value using \textsc{triceratops} \citep{giacalone2020triceratopscode, giacalone2021triceratops}. \textsc{triceratops} estimates the relative probabilities that a given transit signal was produced by a transiting planet or a number of false positive scenarios using a Bayesian analysis. These false positive scenarios include nearby, background, and bound but unresolved eclipsing binaries. For the nine remaining candidates, we obtain a mean FPP of 0.56, and we report the FPP values for our candidates in Table~\ref{tab:followup}. For our full low-mass star sample ($0.088\msun \leq \mstar \leq 0.71\msun$) we calculate an occurrence rate of close-in giant planets of \Nfoccfull.

It is important to note that one must be cautious when using such validation techniques as we use here for giant planets. This is due to the fact that the radii of such planets are very similar to brown dwarfs and very low-mass stars \citep[e.g.][]{mayo2018planetvalidation}. This makes the models used by these validation techniques for giant planets indistinguishable from those for brown dwarfs and low-mass stars. \textsc{triceratops} also includes a prior on planet radius which penalises giant planets with low-mass host stars. These two effects both impact the reliability of the numerical FPP estimates for each candidate, and as such they cannot be used to validate or reject a given individual candidate. However without spectroscopic observations for all candidates they provide the best estimates available to us for the overall false positive probability of our candidates, and therefore they also provide the current best estimate for the occurrence rate of these giant planets. We note that these spectroscopic observations are underway for our candidates.
To investigate the sensitivity of our occurrence rate results on the FPP estimates we consider the cases in which all or none of these candidates are real to set the upper and lower limits of the occurrence rates we calculate. For our full low-mass star sample we calculate such upper and lower limits of \Nfoccfullmax\ and \Nfoccfullmin. We emphasize here that even in the case in which all the currently un-dispositioned candidates are false positives, we have a significantly non-zero occurrence rate for close-in giant planet with low-mass host stars.

\begin{figure*}
    \centering
    \includegraphics[width=0.7\textwidth]{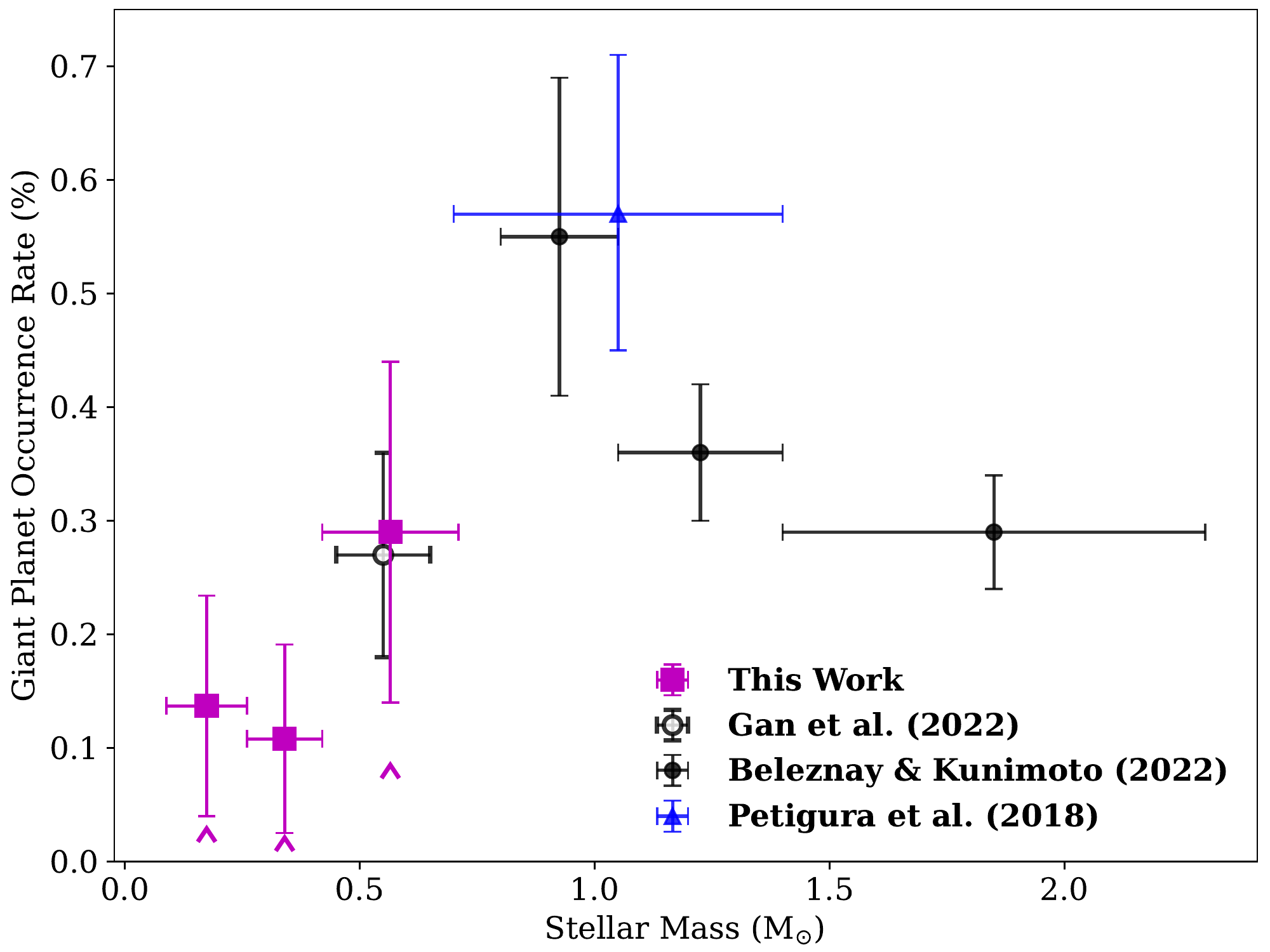}
    \caption{Dependence of the occurrence rate of giant planets with short orbital periods ($P \leq 10\,d$) on the stellar mass of the host star. We compare our results derived in this work (filled magenta squares) to previous TESS results from \citet{beleznaykunimoto2022tesshjoccurrencerates} (filled black circles) and \citet{gan2022tesshjoccurrence} (unfilled black circle). We also compare to the \kepler\ results from \citet{petigura2018cksoccurrencerates} (blue triangle). The x-axis errorbars give the span of the stellar mass ranges used to determine each occurrence rate value. The y-axis errorbars for our results give the standard deviation of the derived Poisson distributions underlying the number of planets we detect. We also plot the 95\,\% lower occurrence rate limits in the case in which none of the remaining nine candidates are real (see text in Section~\ref{sec:occ_rates}), shown by the magenta arrow heads. We note that we have at least one confirmed giant planet in each of our mass ranges, and so we are confident that the occurrence rates in all of the mass ranges are non-zero.}
    \label{fig:focc_vs_stellarmass}
\end{figure*}

The large size of our sample and the stellar mass range it covers allows us to investigate the dependence of giant planet occurrence on stellar mass for our low-mass stars. We separate our stellar sample into three sub-samples with mass ranges defined in order to result in a similar number of stars for each sub-sample, identifying these three similar-size sub-samples as having mass limits of 0.088 - 0.26\,\msun, 0.26 - 0.42\,\msun, and 0.42 - 0.71\,\msun. We calculate occurrence rates for each sub-sample finding values of \Nfocclowmass\ (0.088 - 0.26\,\msun), \Nfoccmedmass\ (0.26 - 0.42\,\msun), and \Nfocchighmass\ (0.42 - 0.71\,\msun). We plot these occurrence rates, along with results from previous studies, in Figure~\ref{fig:focc_vs_stellarmass}, and occurrence rates we calculate in this work are summarised in Table~\ref{tab:occrates}. These occurrence rate measurements are related to Poisson distributions governing the expected number of detected planets. From these distributions we can calculate 95\,\% lower limits for the occurrence rates in each mass range. Doing so we find lower limits of 0.048\,\% (0.088 - 0.26\,\msun), 0.042\,\% (0.26 - 0.42\,\msun), and 0.10\,\% (0.42 - 0.71\,\msun). Each of these three mass ranges also includes at least one of the five confirmed giant planets in our sample, and so in the case where none of the remaining nine candidates are real planets, we calculate occurrence rates of 0.081\,\%, 0.059\,\%, and 0.24\,\%, with 95\,\% lower limits of 0.029\,\%, 0.021\,\%, and 0.085\,\%. Therefore, our results show that a population of close-in giant planets exists even for the lowest range of stellar masses we study.

\begin{figure}
    \centering
    \includegraphics[width=0.95\columnwidth]{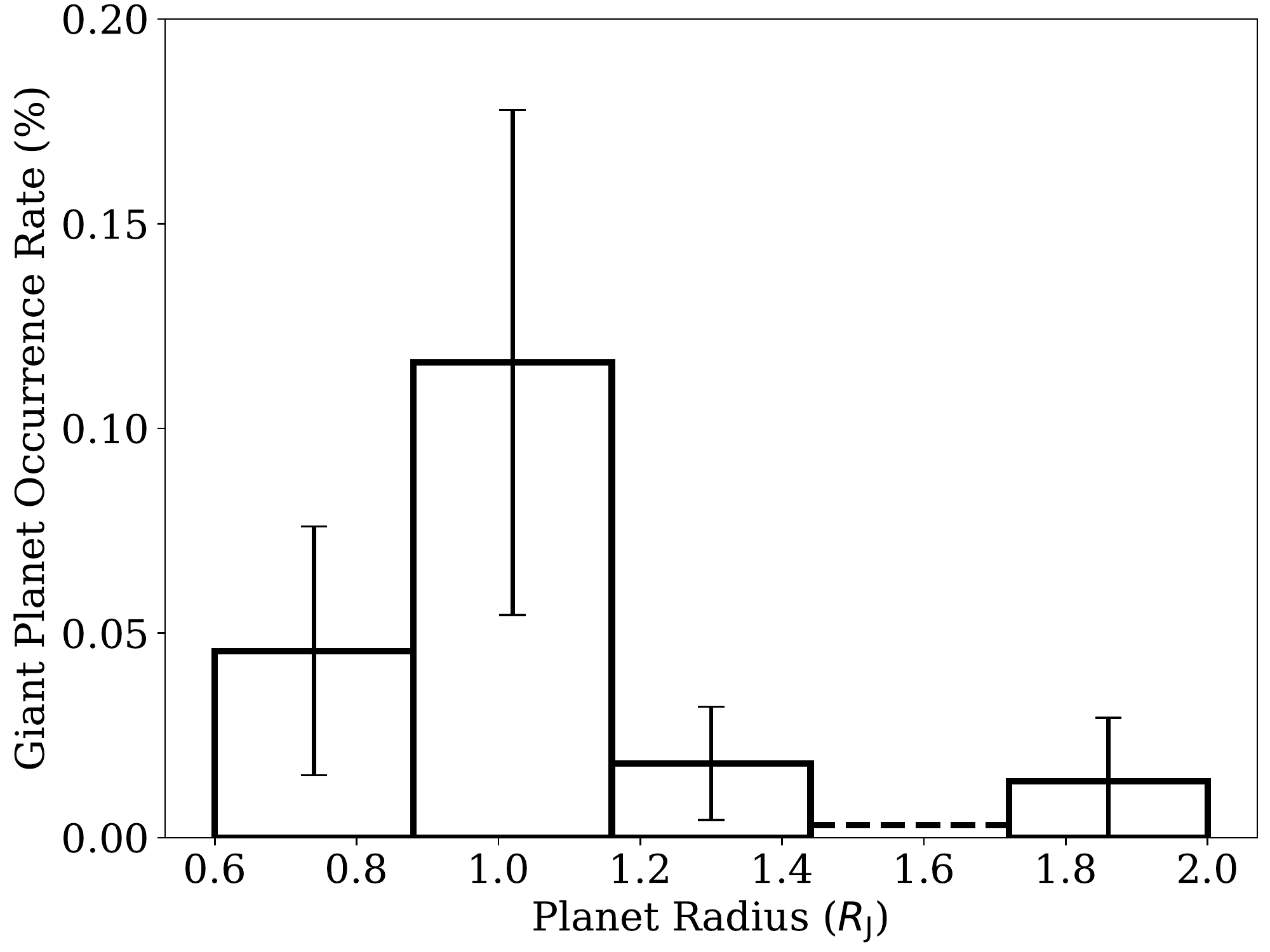}
    \caption{Dependence of our measured occurrence rate of giant planets with short orbital periods ($P \leq 10\,d$) and low-mass host stars on planet radius. Dashed horizontal line represents the 90\% confidence level upper limit for the bin in which there are no planet candidates.}
    \label{fig:focc_vs_rpl}
\end{figure}
\begin{figure}
    \centering
    \includegraphics[width=0.95\columnwidth]{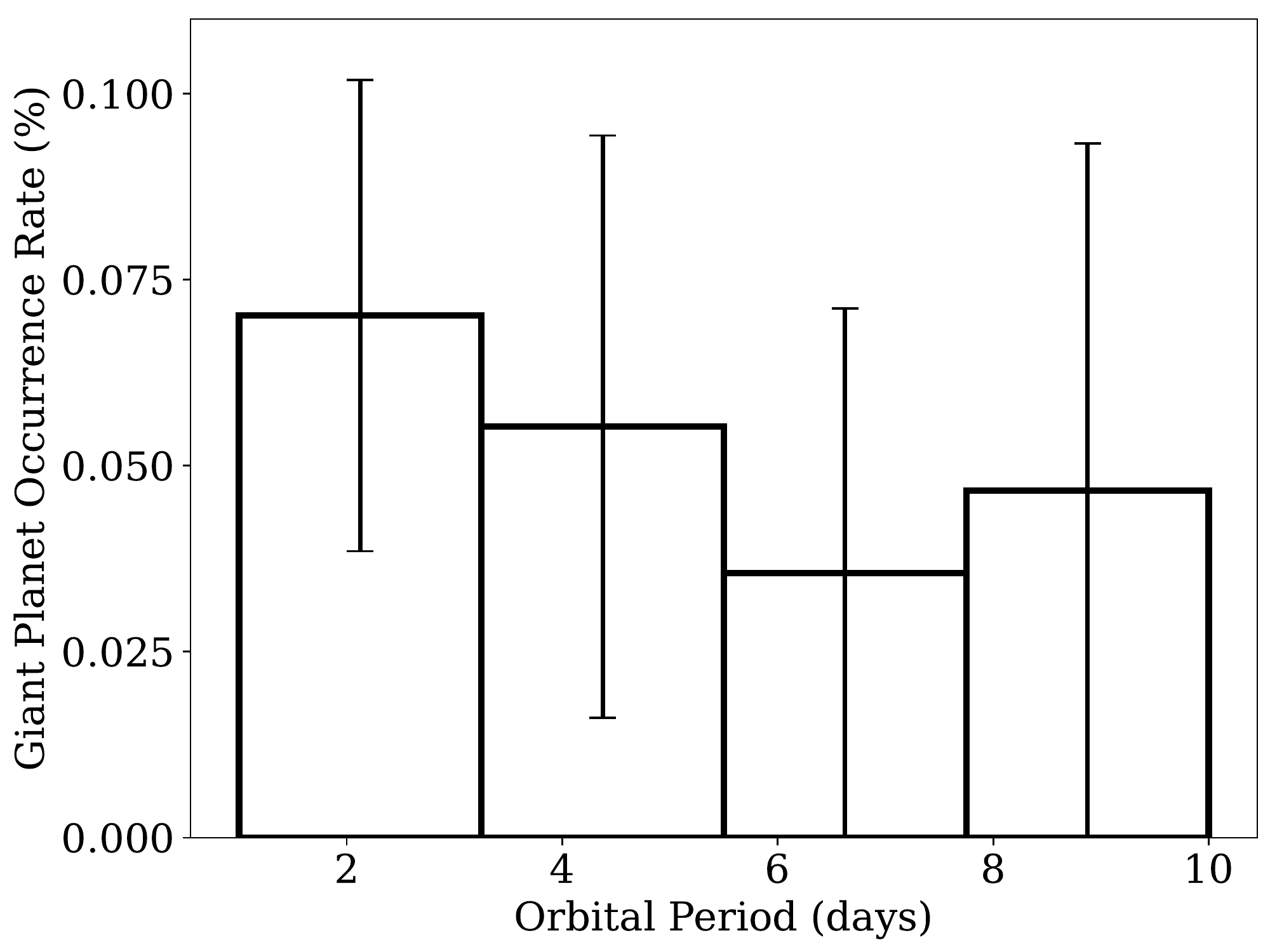}
    \caption{Dependence of our measured occurrence rate of giant planets with short orbital periods ($P \leq 10\,d$) and low-mass host stars on the orbital period of the planet.}
    \label{fig:focc_vs_per}
\end{figure}

The distribution of our candidates in Figure~\ref{fig:detect_eff_full} suggests the possibility of a dependence of the occurrence rate on the planet radius, \rpl. To test whether this distribution is indicative of the underlying population, or simply an effect of the detection efficiency, we calculate the occurrence rate as a function of \rpl. These results are plotted in Figure~\ref{fig:focc_vs_rpl} and we see that the occurrence rates show clear variation with \rpl. We see a peak in occurrence rates for $\rpl \sim 1 \rjup$ and occurrence rates consistent with zero for $\rpl \geq 1.4 \rjup$. Whereas for more massive host stars -- $\mstar \geq 1.2\,\msun$ -- there exist transiting planets with radii $\rpl > 1.5\,\rjup$ (see Figure~\ref{fig:ms_mpl_wcands}).
It has been previously suggested that extreme levels of irradiation of the planet by the star, on the order of $10^5 - 10^6$\,W\,m$^{-2}$, are required to cause the inflation of the planetary atmosphere necessary to reach planetary radii $\gtrsim 1.1 \rjup$ \citep{sestovic2018hjinflation}. A planet in a 3\,d orbit around a 0.45\msun\ star will receive an irradiation of around $5.7 \times 10^4$\,W\,m$^{-2}$. Therefore, it is possible that the derived lack of high radius giant planets for low-mass stars is because these stars are unable to inflate the atmospheres of the planets orbiting them. Previous occurrence rate studies have shown an increase in giant planet occurrence rate for longer orbital periods \citep[e.g.][]{petigura2018cksoccurrencerates}. The distribution of our candidates in Figure~\ref{fig:detect_eff_full} shows that the majority of our candidates have short periods $< 4\,d$. However, the detection efficiencies are also much higher for short orbital periods. Plotting the occurrence rate as a function of orbital period in Figure~\ref{fig:focc_vs_per} we see no clear trend in the occurrence rates. At this stage, the numbers of planets are too low in order to discern a significant trend with orbital period.

\subsection{Comparison with previous works}
Many previous works have measured the occurrence rates of giant planets with varying stellar hosts. In particular, results from the \kepler\ survey -- a sample of F, G, and K dwarf stars -- yielded short period giant planet occurrence rates of $0.43\pm0.05$\% \citep[$\rpl = 6 - 22 \rearth$; $P = 0.8 - 10$\,d]{fressin2013kepleroccurrences} and $0.57^{+0.14}_{-0.12}$\% \citep[$\rpl = 8 - 24 \rjup$; $P = 1 - 10$\,d]{petigura2018cksoccurrencerates}. The occurrence rates we derive here for low-mass stars are significantly lower ($> 1\sigma$) than these \kepler\ results. This is the case even when we consider the upper limit scenario in which all of our candidates are real. More recently, \citet{zhou2019tesshjs} used results from the first seven TESS sectors to study the occurrence rate of hot Jupiters with host stars with varying masses, deriving values of $0.71 \pm 0.31$\% ($0.8 - 1.05\,\msun$), $0.43 \pm 0.15$\% ($1.05 - 1.4\,\msun$), and $0.26 \pm 0.11$\% ($1.4 - 2.3\,\msun$). \citet{beleznaykunimoto2022tesshjoccurrencerates} built upon this early work using photometry from the first two years of the TESS mission to refine these giant planet occurrence rates to $0.55\pm0.14$\% ($0.8 - 1.05\,\msun$), $0.36\pm0.06$\% ($1.05 - 1.4\,\msun$), and $0.29\pm0.05$\% ($1.4 - 2.3\,\msun$). The first TESS giant planet occurrence rate measurement of $0.27\pm0.09$\% for \mdwarf\ host stars (0.45 -- 0.65\,\msun) was provided by \citet{gan2022tesshjoccurrence}.

We compare our results to some of these previous studies in Figure~\ref{fig:focc_vs_stellarmass}. The first clear result from this comparison is that the occurrence rates of short orbital period giant planets decreases with decreasing stellar mass for host stars with $\msun \leq 1\,\msun$. Considering our result from our full sample -- \Nfoccfull\ -- we find it is less than the \citet{petigura2018cksoccurrencerates} result at a level of $2.69\,\sigma$ and less than the \citet{beleznaykunimoto2022tesshjoccurrencerates} $0.8 - 1.05\,\msun$ result at a level of $2.27\,\sigma$. Considering the case in which all nine of the remaining candidates are real planets, our upper limit occurrence rate of \Nfoccfullmax\ is still less than both of these results, at a level of $2.10\,\sigma$ for \citet{petigura2018cksoccurrencerates} and a level of $1.76\,\sigma$ for \citet{beleznaykunimoto2022tesshjoccurrencerates}. Comparing to the \citet{gan2022tesshjoccurrence} results for early \mdwarfs\ we find that our derived result for the $0.42 - 0.71\,\msun$ selection of our stellar sample -- \Nfocchighmass\ -- is fully consistent with their measured occurrence rate for a similar mass range. We note that we have used a different set of TESS FFI light curves than \citet{gan2022tesshjoccurrence}; we have used those produced by the SPOC pipeline while they made use of data from the QLP pipeline \citep{huang2020qlp}. This different set of light curves, along with the use of an independent planet search methodology, allows us to use our results for early \mdwarfs\ to corroborate the \citet{gan2022tesshjoccurrence} findings. The wider stellar mass range of our sample also allows us to measure the occurrence rates for giant planets with much lower mass host stars. If we consider the stars in our sample with $\mstar \leq 0.4 \msun$ we find an occurrence rate of \Nfoccbelowburn, which is less than the \citet{gan2022tesshjoccurrence} result at a level of $1.15\,\sigma$. This is some evidence, if still statistically marginal, that the occurrence rates of giant planets continues to decrease for mid and late \mdwarfs\ compared to early \mdwarfs. Comparing our results from the low and middle mass ranges of our sample to the upper mass range we also find evidence of this decrease for \mdwarfs\ with later spectral types at a similar statistical level. The opposite trend has been observed for small planets, which have been shown to be more common around \mdwarfs\ than more massive host stars \citep[e.g.][]{dressing2015keplermdwarfs, kunimoto2020kepleroccurrence, pinamonti2022hadesmdwarfs}. This difference in occurrence rate variation with mass of the host star could imply a formation efficiency for planets around low-mass stars which favours the formation of multiple small planets over a single giant planet.

\subsection{Implications for giant planet formation}\label{sec:formation_implications}
The dependence of giant planet occurrence rates on the mass of the host star is a clear prediction of the core-accretion planet formation theory \citep[e.g.][]{laughlin2004coreaccretion, burn2021ngpps}. This theory predicts a decrease in occurrence rates for $\mstar < 1\,\msun$ and comparing the results for our full sample to those for solar-like stars from \kepler\ \citep{petigura2018cksoccurrencerates} and TESS \citep{beleznaykunimoto2022tesshjoccurrencerates} we indeed find such a decrease. This agreement in the observed trend with the theoretical prediction suggests that the core-accretion mechanism dominates the formation of these close-in giant planets. The \citet{burn2021ngpps} planet population synthesis, which uses the core-accretion formation theory, also predicts that the formation of giant planets becomes impossible for stars with $\mstar \leq 0.4\,\msun$. Studies of protoplanetary disks have shown their mass to decrease for lower masses of the central star \citep[e.g.][]{andrews2013ppdiskmass, kurtovic2021diskmasses} and therefore for these low-mass stars the predictions are that the disk mass is not sufficient to form a giant planet \citep{pascucci2016ppdiskmasses}. However, our results show that close-in giant planets can and do exist for these low-mass host stars. From our study we derived a giant planet occurrence rate for host stars with $\mstar \leq 0.4\,\msun$ of \Nfoccbelowburn, with a 95\,\% lower limit of 0.05\,\%. In particular, two of our giant planet candidates for which we have obtained spectroscopic follow-up confirmation -- TIC-243641947 and TIC-67512645 -- both have host stars with $\mstar < 0.4 \msun$. So for the case in which we consider that none of our remaining nine giant planet candidates are real, we derive an occurrence rate of 0.074\,\% with a 95\,\% lower limit of 0.026\,\%.

There must be some pathway through which these stars can form giant planets. Firstly, we consider that the current main inhibitor for the predicted formation of these planets is the limits on the mass of the protoplanetary disk. Therefore, if these low-mass stars were capable of supporting much more massive disks than currently expected, even rarely, this would allow for the formation of these giant planets. Observational studies of disk-hosting low-mass stars could allow the occurrence of such massive disks to be constrained. It is important here to note that the dust masses for protoplanetary disks are currently derived from the flux received at millimeter wavelengths, assuming the dust continuum emission is optically thin \citep[e.g.][]{hildebrand1983diskdustmasses, kurtovic2021diskmasses}. If this emission is in fact optically thick, or thicker than currently assumed, then it is possible that we are underestimating the masses of the observed protoplanetary disks. We also note that to date protoplanetary disks have not been observed at a very young age, as at these ages they are still embedded in gas clouds and hard to observe. As such, these disks could begin with a much higher mass than when we can observe them, and so it is possible that their initial masses are sufficient to support giant planet formation.

Another possibility is that these planets did not form through core-accretion, but instead formed by gravitational instability \citep{boss1997gravinstability}. It has been shown that giant planets are capable of forming through this mechanism for host stars with masses as low as 0.1\,\msun\ \citep{boss2006gasgiantformation, mercerstam2020gi}. \citet{mercerstam2020gi} show that the giant planets which form around low-mass stars through gravitational instability have high masses $\gtrsim 2\,\mjup$. 
Spectroscopic follow-up of our candidates is required to measure their masses determine whether they are massive planets that could have formed through gravitational instability.

\section{Conclusions and Future Outlook}
We have presented a systematic search through TESS 30\,minute cadence light curves for transiting gas giant planets orbiting low-mass stars. These systems are predicted to be rare by planet formation theory, and so the aim of this search is to derive a robust occurrence rate for these systems for the first time. We have presented our planet search and candidate vetting pipeline, using which we identified fifteen giant planet candidates from an initial sample of 91,306 low-mass stars. Of these candidates, two are previously known transiting planets \citep{anderson2017wasp107, huang2020toi1130} and through our own preliminary spectroscopic follow-up we have shown three more of our candidates to be giant planets, and one to be an eclipsing binary. Spectroscopic observations over the coming years will be required to uncover the nature of the remaining nine candidates. In addition, seven of our candidates had not been identified as planet candidates prior to this study, including TIC-67512645. Our candidate giant planets have stellar hosts which are in general lower mass than the hosts of currently known transiting giant planets (see Figure~\ref{fig:ms_mpl_wcands}). Therefore, this study is providing strong evidence that the population of transiting giant planets extends to lower stellar mass host stars than previously expected.

We have also performed planet injection-recovery simulations to estimate the detection efficiency of our pipeline. Using these results and our giant planet candidate list we have constrained the occurrence rate of giant planets around low-mass stars, deriving an occurrence rate of \Nfoccfull\ for our low-mass star sample. We also derive occurrence rates for three separate host star mass ranges, calculating values of \Nfocclowmass\ (0.088 -- 0.26\,\msun), \Nfoccmedmass\ (0.26 -- 0.42\,\msun), and \Nfocchighmass\ (0.46 -- 0.71\,\msun), extending our knowledge of giant planet populations to lower stellar mass hosts than previously studied.  The occurrence rate we calculate for our highest mass bin is fully consistent with the results from an independent study focusing on host stars with masses in the range 0.45 -- 0.65\,\msun\ \citep{gan2022tesshjoccurrence}. Comparing with occurrence rate studies for higher mass host stars \citep[e.g.][]{fressin2013kepleroccurrences, petigura2018cksoccurrencerates, beleznaykunimoto2022tesshjoccurrencerates} we demonstrate that giant planets are less common around low-mass stars than solar-type stars, as predicted by the core-accretion planet formation theory \citep[e.g][]{laughlin2004coreaccretion, burn2021ngpps}. Our results provide some strong early evidence that giant planets are even less common around late M-dwarfs than early M-dwarfs but that giant planets can exist with host stars as low mass as $0.2 - 0.3 \msun$. It has previously been asserted that lower protoplanetary disk masses \citep[e.g.][]{pascucci2016ppdiskmasses, kurtovic2021diskmasses} and longer Keplerian timescales \citep[e.g.][]{laughlin2004coreaccretion} completely inhibits the formation of giant planets around stars with masses as low as this. Therefore, while our results for the higher mass stars in our sample are consistent with core-accretion, our results for the lower mass stars in our sample -- $\mstar \leq 0.4\,\msun$ -- present some conflict with the current understanding of how giant planets form. 
Further observations, both to measure the masses of our candidates and study the masses of protoplanetary disks around low-mass stars, will help determine how these planets formed.

The constraints we have placed on giant planet occurrence are currently limited by the fact that two thirds of our candidates are unconfirmed, and they will be significantly strengthened by the confirmation of these candidates. The work to obtain these confirmations is underway and over the next few years we hope to be able to improve our constraints on the occurrence rates as the follow-up effort continues. We also provide full details on our candidates so that the wider community is able to assist in this effort. We note that some of the nine as yet unconfirmed candidates have host stars with $\mstar < 0.4 \msun$. Therefore the new generation of stabilised spectrographs operating at (near-)infrared wavelengths, such as NIRPS \citep{bouchy2017nirps} or SPIRou \citep{thibault2012spirou}, can play a major role in refining the occurrence rates of these systems.

During the fitting analysis outlined in Section~\ref{sec:transit_fitting} we realised the limitations of the long cadence (30\,minute) observations in characterising the true shape of the transit signal. These transit shapes are incredibly important for giant planets transiting small radius stars because there is a much higher probability of the planet crossing the limb. Moving the FFIs to higher cadence - 10\,minutes in the first extended mission and then 200\,seconds in the second extended mission\footnote{\url{https://heasarc.gsfc.nasa.gov/docs/tess/second-extended.html}} - will greatly improve our ability to classify giant planets transiting low-mass stars.

\section*{Acknowledgements}
The contributions at the Mullard Space Science Laboratory by E.M.B. and V.V.E. have been supported by STFC through the consolidated grant ST/W001136/1. We thank Paola Pinilla for useful discussions regarding the planet formation interpretation of our results discussed in Section~\ref{sec:formation_implications}. This paper includes data collected with the TESS mission obtained from the MAST data archive at the Space Telescope Science Institute (STScI). Funding for the TESS mission is provided by the NASA Explorer Program. STScI is operated by the Association of Universities for Research in Astronomy, Ibc., under NASA contract NAS 526555.

\section*{Data Availability}
This work made use of the publicly available TESS-SPOC 30\,minute cadence full-frame-image light curves \citep{caldwell2020tessspoc}. These are available as a High Level Science Product from the Mikulski Archive for Space Telescopes here: \url{https://archive.stsci.edu/hlsp/tess-spoc}.



\bibliographystyle{mnras}
\bibliography{paper} 




\appendix

\section{Stellar parameters for candidate systems}
\begin{table*}
    \centering
        \begin{tabular}{| m{2.3cm} | m{4cm} m{4cm} m{4cm} |}
    \hline
     & & & \\
     & \textbf{TIC-95112238} & \textbf{TIC-311555090} & \textbf{TIC-46432937} \\
     & & & \\
     \textit{TESS} (mag) & $12.2745\pm0.0073$ & $14.8993\pm0.0077$ & $12.3718\pm0.0074$ \\
     \textit{Gaia G} (mag) & $13.5155\pm0.0006$ & $16.2222\pm0.0009$ & $13.433\pm0.0004$ \\
     \textit{Gaia}\,$G_{\rm Bp}$\,(mag) & $14.9506\pm0.0022$ & $17.8488\pm0.0076$ & $14.5196\pm0.0025$ \\
     \textit{Gaia}\,$G_{\rm Rp}$\,(mag) & $12.3406\pm0.0012$ & $14.9837\pm0.0027$ & $12.3996\pm0.0015$ \\
     \textit{J} (mag) & $10.657\pm0.021$ & $13.236\pm0.023$ & $11.011\pm0.022$  \\
     \textit{H} (mag) & $10.063\pm0.024$ & $12.717\pm0.025$ & $10.427\pm0.023$ \\
     \textit{K} (mag) & $9.807\pm0.02$ & $12.468\pm0.023$ & $10.195\pm0.02$  \\
     & & & \\
     & \textbf{TIC-335590096} & \textbf{TIC-406717909} & \textbf{TIC-202468443} \\
     & & & \\
     \textit{TESS} (mag) & $13.7726\pm0.0079$  & $13.5451\pm0.0080$  & $14.5982\pm0.0074$\\
     \textit{Gaia G} (mag) &  $15.1091\pm0.0013$ &  $14.8626\pm0.0009$ & $15.9658\pm0.0009$\\
     \textit{Gaia}\,$G_{\rm Bp}$\,(mag) &  $16.7462\pm0.0062$ & $16.454\pm0.0063$  & $17.7402\pm0.0072$\\
     \textit{Gaia}\,$G_{\rm Rp}$\,(mag) &  $13.8593\pm0.0035$ &$13.6273\pm0.0037$  & $14.6966\pm0.0014$\\
     \textit{J} (mag) & $12.056\pm0.022$  & $11.897\pm0.021$  & $12.818\pm0.021$\\
     \textit{H} (mag) &  $11.431\pm0.026$ & $11.3\pm0.027$  & $12.218\pm0.023$\\
     \textit{K} (mag) & $11.175\pm0.026$  & $11.079\pm0.023$ & $11.965\pm0.022$\\
     & & & \\
     & \textbf{TIC-165227846} & \textbf{TIC-60910638} & \textbf{TIC-73692250} \\
     & & & \\
     \textit{TESS} (mag) & $13.4581\pm0.0074$&  $13.0687\pm0.0077$ & $13.9856\pm0.0077$ \\
     \textit{Gaia G} (mag) & $14.8075\pm0.0005$ &$14.3144\pm0.0011$  & $15.3646\pm0.0009$ \\
     \textit{Gaia}\,$G_{\rm Bp}$\,(mag) & $16.5098\pm0.0053$ & $15.7465\pm0.0077$ & $17.1267\pm0.0057$ \\
     \textit{Gaia}\,$G_{\rm Rp}$\,(mag) & $13.5507\pm0.0015$ & $13.1344\pm0.0028$ &  $14.0843\pm0.0028$\\
     \textit{J} (mag) & $11.743\pm0.021$ & $11.487\pm0.02$  & $12.239\pm0.025$ \\
     \textit{H} (mag) & $11.169\pm0.022$ &$10.894\pm0.023$  & $11.681\pm0.025$ \\
     \textit{K} (mag) & $10.885\pm0.019$ & $10.654\pm0.014$ & $11.409\pm0.02$\\
     & & & \\
     & \textbf{TIC-243641947} & \textbf{TIC-77490011} & \textbf{TIC-67512645} \\
     & & & \\
     \textit{TESS} (mag) & $13.2486\pm0.0078$ & $14.1398\pm0.0074$ & $14.9046\pm0.0078$ \\
     \textit{Gaia G} (mag) & $14.4858\pm0.0007$ & $15.4618\pm0.0006$ &  $16.3092\pm0.0012$ \\
     \textit{Gaia}\,$G_{\rm Bp}$\,(mag) & $15.9209\pm0.0023$ & $17.0973\pm0.0036$ & $18.1135\pm0.0234$ \\
     \textit{Gaia}\,$G_{\rm Rp}$\,(mag) & $13.3138\pm0.0032$ & $14.2251\pm0.0015$ &  $15.0078\pm0.0026$ \\
     \textit{J} (mag) &  $11.706\pm0.025$ & $12.479\pm0.021$ & $13.169\pm0.023$ \\
     \textit{H} (mag) &  $11.099\pm0.024$ & $11.930\pm0.023$ & $12.486\pm0.022$ \\
     \textit{K} (mag) &  $10.819\pm0.021$ & $11.622\pm0.019$ & $12.207\pm0.021$ \\
     & & & \\
     & \textbf{TIC-178709444} & \textbf{TIC-429302040} & \textbf{TIC-254113311} \\
     & & & \\
     \textit{TESS} (mag) & $13.6615\pm0.0073$& $10.4180\pm0.0061$ & $10.1429\pm0.0061$ \\
     \textit{Gaia G} (mag) & $14.9535\pm0.0005$& $11.1740\pm0.0009$ & $10.9028\pm0.0008$ \\
     \textit{Gaia}\,$G_{\rm Bp}$\,(mag) & $16.5039\pm0.0031$ & $11.9254\pm0.0012$ & $11.6530\pm0.0028$ \\
     \textit{Gaia}\,$G_{\rm Rp}$\,(mag) & $13.7388\pm0.0014$& $10.3741\pm0.0007$ & $10.0917\pm0.0014$ \\
     \textit{J} (mag) & $11.999\pm0.026$ & $9.378\pm0.021$ & $9.055\pm0.023$ \\
     \textit{H} (mag) &  $11.348\pm0.028$& $8.777\pm0.026$ & $8.493\pm0.059$ \\
     \textit{K} (mag) &  $11.084\pm0.027$ & $8.637\pm0.023$ & $8.351\pm0.033$ \\
     & & & \\
    \hline
    \end{tabular}
    \caption[Stellar magnitudes for our giant planet candidates]{Stellar magnitudes for the host stars of our giant planet candidates. The values are from TICv8 \citep{stassun2019ticv8}, \textit{GAIA} DR2 \citep{GAIA_DR2}, and 2MASS \citep{skrutskie2006twomass}.}
    \label{tab:stellar_mags}
\end{table*}
 
\begin{table*}
    \centering
        \begin{tabular}{| m{2.2cm} | m{4cm} m{4cm} m{4cm} |}
    \hline
    & & & \\
     & \textbf{TIC-95112238} & \textbf{TIC-311555090} & \textbf{TIC-46432937} \\
    & & & \\
     RA (deg) & $90.4837848985143$   & $87.0953906689957$   & $83.8690390629687$   \\
     Dec (deg) &   $-16.8166146303267$   &   $53.491840564709$   &   $-14.59734983288$   \\
     Parallax (mas) &   $12.543\pm0.045$   &   $11.224\pm0.105$   &   $11.016\pm0.025$   \\
     Distance (pc) &   $79.72\pm0.29$   &   $89.09\pm0.84$   &   $90.77\pm0.21$   \\
     \rstar\ (\rsun) &   $0.557\pm0.017$   &   $0.235\pm0.007$   &   $0.539\pm0.016$   \\
     \mstar\ (\msun) &   $0.552\pm0.020$   &   $0.205\pm0.020$   &   $0.535\pm0.020$   \\
     \teff\ (K) &   $3374\pm157$   &   $3247\pm157$   &   $3673\pm157$   \\
     \logg &   $4.6878\pm0.0098$ &   $5.0074\pm0.0158$ &   $4.7037\pm0.0093$ \\
     & & & \\     
     & \textbf{TIC-335590096} & \textbf{TIC-406717909} & \textbf{TIC-202468443} \\
    & & & \\
     RA (deg) & $183.5647447$  & $309.47724827762$ & $229.006270911694$   \\
     Dec (deg) & $-13.17481731$  & $19.5764691725682$ &   $62.7183462414147$   \\
     Parallax (mas) & $12.450\pm0.099$ & $10.122\pm0.046$ &   $10.273\pm0.063$   \\
     Distance (pc) &  $80.32\pm0.64$  & $98.79\pm0.45$ &   $97.35\pm0.59$   \\
     \rstar\ (\rsun) & $0.353\pm0.011$ & $0.428\pm0.013$ &   $0.307\pm0.009$   \\
     \mstar\ (\msun) & $0.338\pm0.021$  & $0.422\pm0.020$ &    $0.285\pm0.020$  \\
     \teff\ (K) & $3237\pm157$  & $3266\pm157$ &   $3162\pm157$   \\
     \logg & $4.87111\pm0.00054$ & $4.8014\pm0.0052$ &   $4.9177\pm0.0045$ \\
     & & & \\
     & \textbf{TIC-165227846} & \textbf{TIC-60910638} & \textbf{TIC-73692250} \\
    & & & \\
     RA (deg) & $178.819448968594$   & $5.68952194396963$   & $40.0813749354698$ \\
     Dec (deg) &   $-40.1491860945041$   &   $22.9455977613927$   & $35.4105323079258$ \\
     Parallax (mas) &   $16.038\pm0.074$   &   $10.909\pm0.073$   & $11.749\pm0.087$ \\
     Distance (pc) &   $62.35\pm0.29$   &   $91.67\pm0.62$   & $85.11\pm0.63$ \\
     \rstar\ (\rsun) &   $0.321\pm0.010$   &   $0.467\pm0.014$   & $0.339\pm0.010$ \\
     \mstar\ (\msun) &   $0.300\pm0.020$   &   $0.465\pm0.020$   & $0.322\pm0.020$ \\
     \teff\ (K) &   $3202\pm157$   &   $3373\pm157$   & $3163\pm157$ \\
     \logg &   $4.9034\pm0.0032$ &   $4.7660\pm0.0068$  & $4.8847\pm0.0011$ \\
     & & & \\
     & \textbf{TIC-243641947} & \textbf{TIC-77490011} & \textbf{TIC-67512645} \\
    & & & \\
     RA (deg) &  $207.4749061$  & $42.8241708040705$   & $173.4697891$ \\
     Dec (deg) & $-46.06623829$ &   $30.2817681183302$   & $12.45099531$ \\
     Parallax (mas) & $13.707\pm0.099$  &   $16.749\pm0.068$   & $13.52\pm0.0968$  \\
     Distance (pc) &  $72.96\pm0.53$ &   $59.70\pm0.24$   &  $73.96\pm0.53$  \\
     \rstar\ (\rsun) & $0.372\pm0.011$ &   $0.233\pm0.007$   & $0.2221\pm0.0068$ \\
     \mstar\ (\msun) &  $0.360\pm0.020$  &   $0.203\pm0.020$   &  $0.191\pm0.020$\\
     \teff\ (K) & $3375\pm157$  &   $3244\pm157$   & $3133\pm157$ \\
     \logg & $4.8525\pm0.0018$  &   $5.0103\pm0.0171$ &  $5.027\pm0.019$  \\
     & & & \\
     & \textbf{TIC-178709444} & \textbf{TIC-429302040} & \textbf{TIC-254113311} \\
    & & & \\
     RA (deg) & $166.0757629$ & $188.386850399595$   & $286.375931663504$   \\
     Dec (deg) &  $-47.82139085$  &   $-10.1461733018634$   &   $-41.4375228598884$   \\
     Parallax (mas) &  $10.129\pm0.044$ &   $15.418\pm0.062$   &  $17.136\pm0.050$    \\
     Distance (pc) &$98.72\pm0.43$  &   $64.86\pm0.26$   &   $58.36\pm0.17$   \\
     \rstar\ (\rsun) &  $0.427\pm0.013$ &   $0.720\pm0.065$   &    $0.742\pm0.068$  \\
     \mstar\ (\msun) & $0.421\pm0.021$ &   $0.660\pm0.080$   &   $0.660\pm0.076$   \\
     \teff\ (K) &  $3296\pm157$ &   $4251\pm121$   &   $4236\pm122$   \\
     \logg &  $4.8023\pm0.0053$ &   $4.5435\pm0.1034$ &   $4.5169\pm0.1023$ \\
     & & & \\
     \hline
    \end{tabular}

    \caption[Stellar parameters the host stars of our giant planet candidates]{Stellar parameters for the host stars of our giant planet candidates. The parameters have been taken from the TICv8 \citep{stassun2019ticv8}.}
    \label{tab:stellar_params}
\end{table*}
Here we provide details on the stellar parameters of the host stars for our candidate giant planet systems.


\bsp	
\label{lastpage}
\end{document}